\documentclass[12pt]{iopart}
\usepackage{graphicx}

\usepackage[dvipsnames]{xcolor}
\begin{document}

\title[On the edge turbulence..]{On the edge turbulence in a DTT-like tokamak plasma}

\author{F. Cianfrani$^1$ \& G. Montani$^{1,2}$}

\address{$^1$ ENEA, Nuclear Department, C. R. Frascati, Via E. Fermi 45, 00044 Frascati (Roma), Italy}
\address{$^2$ Physics Department, “Sapienza” University of Rome, P.le Aldo Moro 5, 00185 Roma, Italy}
\eads{\mailto{francesco.cianfrani@enea.it}, \mailto{giovanni.montani@enea.it}}
\vspace{10pt}
\begin{indented}
\item[]April 2024
\end{indented}

\begin{abstract}
Turbulent transport provides the main contribution to particle and energy losses in tokamak plasmas, which control is of paramount importance for forthcoming reactors such as the Divertor-Tokamak-Test (DTT) facility under construction at ENEA Frascati. In this work we investigate the characteristic features of drift turbulence at the plasma edge through 3D electro-static fluid simulations. We outline the crucial role of the diffusion coefficient for the emerging turbulent spectra and for the excitation of vortex structures or zonal flows. Moreover, the impact of adding a poloidal magnetic component is discussed considering also a radial shear, and the emergence of anisotropic spectral features is emphasized. The analysis is extended to the case with Dirichlet boundary conditions along the radial direction, instead of the periodic ones usually employed in such kind of analyses.          
\end{abstract}

%
%
%
%
%

\section{Introduction}
The confinement of particles and energy in the core of fusion plasmas is the main challenge for nuclear fusion research \cite{Ding:24}. The realization of a spatial region with closed magnetic field lines (core) is the key aspect of tokamak reactors, in which a toroidal magnetic field is applied via external coils and a poloidal component is generated inductively through an electric discharge. The plasma core dynamics is characterized by violent instabilities and strong non-linearities, generating turbulent outward transport reducing confinement that can only be described via extremely costly (gyrokinetic) numerical simulations \cite{Mantica:20,Mishchenko:23}. 
The core is surrounded by the scrape-off-layer (SOL) in which the magnetic field lines open and can be tailored to form an X-point and drive most of the particle fluxes in the so-called divertor region where they impact some properly designed targets capable of sustaining large heat fluxes \cite{Pitcher:97}. In order to test the divertor design for ITER and DEMO, the Divertor-Tokamak-Test facility (DTT) \cite{DTT,DTT2} is being realized at ENEA laboratories in Frascati. 

Physical conditions change moving from the core to SOL with increasing collisionality and decreasing energy/temperature/density, such that Braginski fluid models \cite{Braginskii:1965} with drift ordering \cite{Simakov:2004} are expected to be viable and are commonly used by the most advanced global codes simulating edge plasma turbulence (as for instance BOUT++\cite{Shanahan:14}, GBS \cite{Halpern:16}, TOKAMAK3X \cite{Tamain:16} and GRILLIX \cite{Stegmeier:18}, see Ref.\cite{Schwander:24} for a recent review). These codes are aimed at predicting the profiles of the main physical quantities (density, temperature, electrostatic and magnetic potentials) measured in experiments accounting for the complicated wall features and magnetic geometries of real machines (see for instance \cite{Oliveira:22}) and recently including also neutrals \cite{Giacomin:22,Mancini:24,Quadri:24} to improve code validation in the divertor region.     

The focus of the present work is complementary to these analyses: we want to provide an enhancement in the physical comprehension of tokamkak edge plasma turbulence, by exploiting the impact of the adopted parameters and of the (X-point) magnetic geometry in a simplified context that could help interpreting the results of more comprehensive codes (as for instance turbulence suppression near the X-point \cite{Stegmeier:19}, poloidal asymmetry \cite{Bufferand:21}, generation of sheared flows/electric fields \cite{Zholobenko:21}, global vs local effects..). 

The standard model for edge drift turbulence is the Hasegawa-Wakatani (HW) \cite{HW1,HW2,HW3} (see also Ref.\cite{Hasegawa:18} for an historical review) system of equations, in which vorticity is subject to the Euler equation and density is passively advected on the $\mathrm{E\times B}$ velocity field while the drift term couples the two equations through second order parallel gradients. Because of the anisotropy due to the background magnetic field 2D HW models are usually considered with an effective drift coupling obtained by replacing second parallel derivatives with the so-called adiabatic parameter.    
The seminal work \cite{Biskamp:95} outlined the different behavior of 3D simulations with respect to effective 2D models. In 3D turbulence is sustained by the nonlinear transfer between $n=0$ and $n\neq 0$ modes, the latter decaying in time such that turbulence quench and a poloidal flow tends to form (self-regulation). The analysis in Ref.\cite{Biskamp:95} was performed in SLAB and hyperdiffusive and hyperviscous dissipation terms were adopted. 
A more complete model for edge tokamak plasma was investigated in Ref.\cite{Scott:02}, including electro-magnetic fluctuations, interchange instability and sheared magnetic geometry. The main achievement was the recognition that drift turbulence is a self-sustained process \cite{Scott:90}, {\it i.e.} that the presence of linear instability is not much relevant for the nonlinear regime, as mode saturated levels and energy transport are mostly determined by nonlinear mechanisms rather than by the broadening of linear instabilities (an assumption made for instance in the mixing length argument). It is was also discussed the different role of nonlinear advection for vorticity and for pressure, with the latter providing sub-grid dissipation and the former redistribution to all modes sustaining turbulence.  

Self-sustainance was the motivation for the reduced model of edge turbulence described in Refs.\cite{Montani:22,Montani:23,Carlevaro:23}, in which the linear drive due to background pressure gradients was neglected and the HW system was reduced to a single equation for the electro-static potential only.  
    
The generation of toroidally symmetric poloidal (zonal) flows \cite{Diamond:05,Diamond:11} has a crucial impact on turbulence. In particular, sheared flows break up short-scale vortex structures and suppress turbulence. This is accompanied by the formation of a sheared radial electric field and a transport barrier \cite{Burrell:20,Vermare:22}. In general, the fate of turbulence is affected by spectral symmetry. In fact, a symmetry breaking in the spectral correlator of collisional drift wave turbulence is known to produce finite residual (non-diffusive) Reinolds stress component whose divergence is responsible for the generation of momentum flows quenching turbulence \cite{Hidalgo:06,Long:19}.  		
		
In this work, we study HW system of equations by performing fast 3D simulations in Fourier space and comparing the results varying the main linear parameters, namely the background pressure gradient length scale $\ell_0$ and diffusivity $\chi_\perp$. Particular attention is devoted to the dependence on the turbulence features on the value of the particle diffusion coefficient, {\it i.e.} Braginskii \cite{Braginskii:1965} and neoclassical \cite{Connor:73} setups are considered. We will outline the different character of the resulting turbulent scenarios, in particular the achievement of toroidally symmetric and of poloidally symmetric configurations (zonal flows), the profiles of the resulting 2D spectra in the perpendicular plane and the development of anisotropic profiles, the role of toroidal transitions ($n\neq 0\rightarrow n=0$) in establishing the saturated energy spectra. In order to understand the impact of the magnetic geometry we consider three different magnetic configurations. The SLAB case in which the magnetic field is purely toroidal will set the benchmark scenarios for drift turbulence. By the addition of a constant poloidal component (limiter-like configuration), we will discuss the impact of a poloidal drift component on such scenarios, that is to provide anisotropic spectra and to modify the saturated behavior of toroidal transitions. The stabilizing role of shear \cite{Kendl:03} will finally be addressed by considering a radially sheared polodial magnetic component, that will still lead to anisotropic spectra whose profiles will yet approach those in SLAB.  
 
We will also consider nontrivial boundary conditions with respect to fully periodic ones inherent to the analysis in Fourier space. In fact, we will show how vanishing Dirichlet boundary conditions can be realized through a formulation in terms of the sine Fourier transform. This procedure will be implemented along the radial direction and will produce a further source of spectral anisotropy, and partial turbulence quench, that can be understood in terms of symmetry breaking due to the different boundary conditions for the radial coordinate and the poloidal angle.      

The main merit of this manuscript is in outlining that in the neoclassical regime of particle transport the basic drift coupling mechanism underlying self-sustained turbulence is suppressed in favor of the formation of toroidally and poloidally symmetric structures. 
 
The organization of the work is as follows. In section \ref{sec2} the basic equations for the edge turbulence model are presented, while in subsection \ref{sec2.1} the numerical integration scheme is described. The simulation results are shown in section \ref{sec3}, with the different considered magnetic geometries discussed in next subsections: SLAB in subsection \ref{sec3.1}, constant and radially sheared poloidal magnetic components in subsections \ref{sec3.2} and \ref{sec3.3}, respectively. In section \ref{sec4} concluding remarks follow. 

\section{Basic equations and methodology}\label{sec2}
Drift turbulence in the edge plasma region can be described in the low frequency limit ($\omega<<\Omega_i$, $\Omega_i$ being ion cyclotron frequency) using the fluid description with drift-ordering approximation. Assuming charge neutrality, equal electron and ion temperatures $T_e=T_i=T$ and a background pressure $p_0$, such that the total pressure $p=p_0+\tilde{p}$, the variables are the electro-static potential $\phi$, the perturbed pressure $\tilde{p}$, the magnetic potential $\psi$ and the parallel fluid velocity $v_\parallel$. Discarding $v_\parallel$ and parallel diffusivity, the basic equations can be derived from charge neutrality, energy balance and generalized Ohm's law resulting in \cite{Beyer:98,Montani:23}
\begin{equation}
\left\{\begin{array}{c} 
\partial_t \Delta_\bot \phi + \{\phi,\Delta_\bot \phi\} = B\,\nabla_\parallel\left(\Delta_\bot\psi\right)  + \mu \,\Delta_\bot^2\phi\hspace{.5cm}\\
\partial_t \tilde{p} + \{\phi,\tilde{p}\} = -\hat{k}\,\delta_{y} \phi + D\,\nabla_\parallel \left(\Delta_\bot\psi\right) + {\bf{\chi_\bot}}\, \Delta_\bot\tilde{p}  \\
\partial_t \psi =  F\,\nabla_\parallel (\tilde{p}-\phi) + \eta\,\Delta_\bot\psi  
\end{array}
\right.\label{EM}
\end{equation}
where ion diamagnetic velocity and magnetic (curvature and gradient) drifts have been neglected. Parallel gradient $\nabla_\parallel$ appears in those terms accounting for drift response on the right-hand sides of the equations above and they are 
defined with respect to the magnetic geometry via the corresponding unit vector $\hat{b}=\vec{B}/B$
\begin{equation}
\nabla_\parallel= \hat{b}_x \partial_x + \hat{b}_y \partial_y + \hat{b}_z \partial_z\,,
\end{equation}
with $x$, $y$ and $z$ local Cartesian coordinates for the radial, poloidal and toroidal direction, respectively. The perpendicular Laplacian reads 
\begin{equation}
\Delta_\bot=\Delta-(\nabla_\parallel)^2 \,,
\end{equation}
and it determines vorticity $\Delta_\bot \phi$ and the dissipative contributions.
 
The second terms on the left-hand side of the first two equations in (\ref{EM}) provide the advection by $\mathrm{E\times B}$ drift and are written in terms of the Poisson brackets
\begin{equation}
\{f,g\}=\delta_{x} f\,\delta_{y} g - \delta_{x} g\,\delta_{y} f  
\end{equation}
$\delta_{x}$ and $\delta_{y}$ being the corresponding derivative operators projected onto the perpendicular plane to the magnetic field. The first term on the right-hand side of the equation for $\tilde{p}$ is due to the background pressure gradients since $\hat{k}=\delta_{x} p_0$ and it provides a free-energy source for turbulence. In what follows, a linear background pressure profile is assumed, such that $\hat{k}=\delta_{x} p_0\sim \partial_x p_0=1/\ell_0$ can be taken as a constant with $1/\ell_0$ specifying the magnitude of $p_0$ gradient.  

The last terms on the right-hand side of Eqs.(\ref{EM}) are dissipative. In what follows, the corresponding parameters, namely viscosity $\mu$, diffusivity $\chi_\bot$ and resistivity $\eta$, are all fixed according with Braginskii values, except for $\chi_\bot$ for which also a value ten times larger is considered representing the neoclassical case $\chi_\mathrm{neocl}=10\chi_\mathrm{class}$. 

In the electro-static limit, one can assume $\partial_t\psi\sim 0$ and compute $J_\parallel=\Delta_\bot\psi$ from the generalized Ohm's law so getting the reduced set of equations
\begin{equation}
\left\{\begin{array}{c} 
\partial_t \Delta_\bot \phi + \{\phi,\Delta_\bot \phi\} = \displaystyle\frac{BF}{\eta}\,\nabla_\parallel^2 (\tilde{p}-\phi)  + \mu \,\Delta_\bot^2\phi\hspace{.5cm}\\
\partial_t \tilde{p} + \{\phi,\tilde{p}\} = -\hat{k}\,\delta_{y} \phi + \displaystyle\frac{DF}{\eta}\,\nabla_\parallel^2 (\tilde{p}-\phi) + \chi_\bot\, \Delta_\bot\tilde{p}  
\end{array}
\right.\label{ES}
\end{equation}

A typical DTT-like scenario is considered having \cite{DTT}
\begin{equation}
n_0=5\cdot 10^{-13}cm^{-3}\quad T= 100eV,\quad B=3T\quad R=2.14m\quad a=68cm.
\end{equation}
$n_0$ being the background number density, $R$ and $a$ the major and minor tokamak radius, respectively. The corresponding Braginskii dissipative parameter $\chi_{\bot}$ read  
\begin{equation}  
\chi_{\bot}=\chi_\mathrm{class}=110\,cm^2/s\,.
\end{equation}

The electro-static and magnetic potential are normalized with $e/K_BT$, $K_B$ being the Botzmann constant, and pressure with $n_0K_BT$. 
A fully-toroidal local polodial section in the form of a 2cm side square is simulated in what follows neglecting any curvature effect. Cartesian coordinates are rescaled as $x,y\rightarrow x/L_\perp,y/L_\perp$ and $z\rightarrow z/L_\parallel$ such that rescaled coordinates range in the interval [0,10].   

\subsection{Numerical scheme}\label{sec2.1}

The system of equations (\ref{ES}) is solved using the explicit 4$^\mathrm{th}$ order Runge-Kutta method in Fourier space (FS). At each iteration $\tilde{p}$ and vorticity are advanced, while the electrostatic potential is computed by inverting $\Delta_\perp$ in FS. All the nonlinear terms are computed in real space and pulled-back in FS (pseudospectral method). The simulations are run in python that contains the \emph{scipy} package providing an efficient implementation of the Fast Fourier Transforms (FFT) algorithm to move quickly back and forth to real space. 

\begin{figure}[h]%
\centering
\includegraphics[height=7cm,width=.7\textwidth]{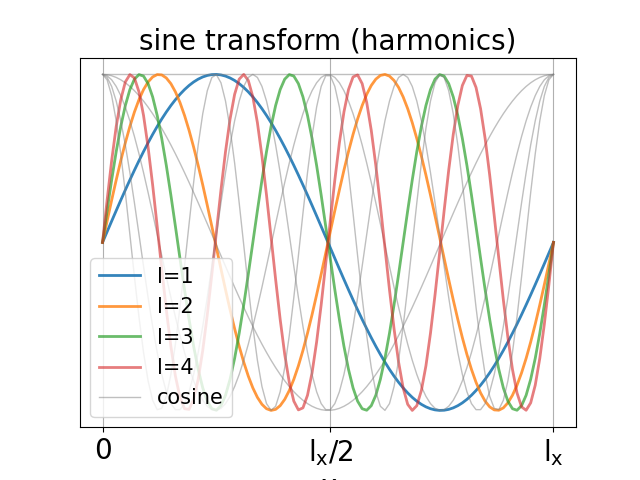}
\caption{The first four sine (blue, orange, green and red lines) and cosine (light grey lines) harmonics in the radial simulation domain. Sine functions vanish at the boundary $x=0,\ell_x$ and at the midpoint $x=\ell_x/2$. The profile in the second half is a specular copy of that in the first half with the mirror placed at the midpoint.}
\label{fig_sineh}
\end{figure}

The description in FS naturally provides periodic boundary conditions (BC) for all the coordinates. However, we also implement vanishing Dirichlet BC in the $x$ radial direction. This is done by restricting to $x$ sine Fourier transform, such that the resulting field profiles are skew-symmetric under reflection along the midpoint $\ell_x/2$ of the simulation domain and vanish in $0$ and $\ell_x/2$ (see Fig.\ref{fig_sineh}). All the background quantities, namely the background pressure gradient and the magnetic configuration, are taken symmetric under the same kind of reflection so preserving fields skew-symmetry along $x$ in time and obtaining a solution of the considered problem with vanishing Dirichlet BC on half of the simulation domain. 

\begin{figure}[h]%
\centering
\includegraphics[width=\textwidth]{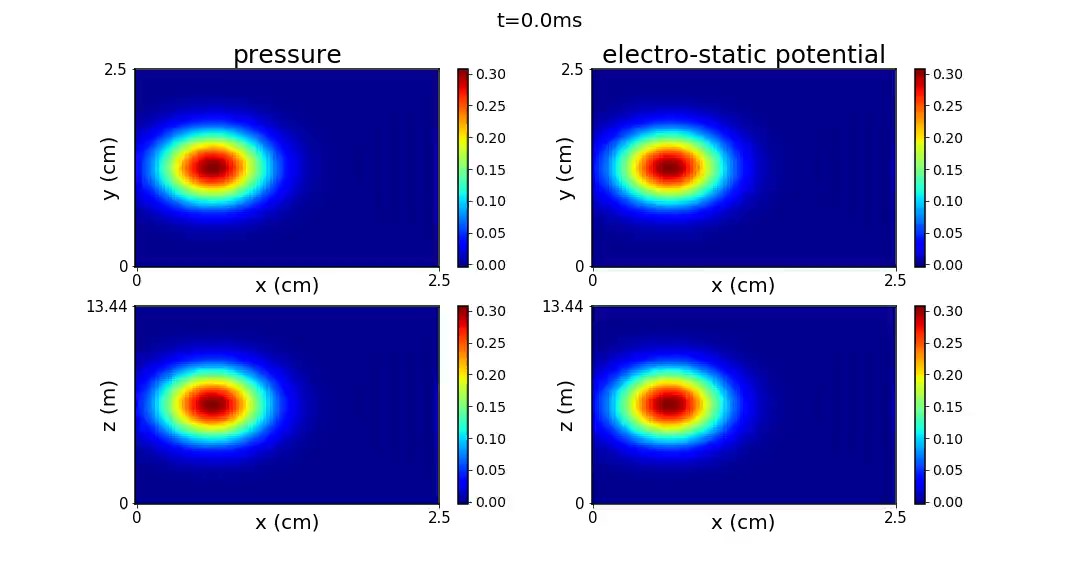}
\caption{The 3D Gaussian functions taken as initial conditions of the simulations: the first row contains the poloidal sections, $(x,y)$ plane, of normalized $\tilde{p}$ (left) and $\phi$ (right) while the second row is for the radial-toroidal $(x,z)$ plane. The corresponding colorbars are shown next to each plot.}
\label{fig_init}
\end{figure}

The chosen initial conditions are 3D Gaussian wave packets centered inside the simulation domain as shown in Fig.\ref{fig_init} with $\phi=\tilde{p}$ to minimize initial drift response. The initial values are such that turbulence is self-sustained \cite{Scott:90}, {\it i.e.} it is not seed by a linear unstable phase. We consider two cases for background pressure with $\ell_0=2\mathrm{cm}$ and $\ell_0=10\mathrm{cm}$ denoted as strong and weak gradients, respectively. Ignoring the role of the magnetic geometry ({\it i.e.} the stabilizing contribution of shear), a linear unstable spectral region is present in the former case, while all the modes are stable in the latter. In both cases we will see turbulent behavior. Linear spectral analysis can be done analytically (see for instance \cite{Numata:07}). A linear phase has here been found by reducing the magnitude of the initial condition and it has been used for benchmarking.  

Two value for diffusivity are discussed: Braginskii (classical case, $\chi_\mathrm{class}$) and ten times larger, that is representative of the neo-classical value $\chi_{\mathrm{neocl}}=10\chi_\mathrm{class}$. We will emphasize the different behavior of turbulent spectra in such two cases. The value of perpendicular diffusivity in turbulent tokamak edge plasma simulations is usually fixed in order to reproduce expected features (pressure and density profiles) or to avoid numerical instabilities \cite{Scott:07} and not according to a self-consistent procedure. This is due to the expectation that turbulent transport dominates diffusivity so that the bare diffusivity value can be ignored. Here we will outline how the resulting turbulence is strongly affected by the choice of classical or neoclassical diffusivity, suggesting that the chosen value of bare diffusivity may indeed play a role in determining the fate of turbulence and the emerging particle and energy transport. 

The results for three magnetic polodial configurations in addition to the constant toroidal component are presented in the next sections: vanishing (SLAB), constant and radially sheared fields. In the first two cases all $\hat{b}$ components are constant (no shear) and $\phi$ is analytically derived from vorticity. Advection terms contain the products of two field derivatives that are computed in real space and pulled back to FS with FFT. Aliasing error is avoided by explicit zero padding for one third of the modes (implicit padding can be used for MPI implementations that are not considered here \cite{Bowman:2011}). 

In the last case magnetic shear is present. The Laplacian and its inverse are numerically pre-computed before starting the iteration. The drift terms contain the expression $\nabla_\parallel(\tilde{p}-\phi) =\hat{b}\cdot\vec{\nabla}(\tilde{p}-\phi)$  and since $\hat{b}$ is not constant they are treated just like nonlinear advection in those cases without shear, {\it i.e.} $(\tilde{p}-\phi)$ gradients are mapped to real space, multiplied times $\hat{b}$ components and finally pulled back to FS. The advection terms are now given by the product two fields and one geometric function, obtained by expanding $\delta_{x}$ and $\delta_{y}$ in $x$, $y$, $z$ derivatives, resulting in a stronger mode restriction to avoid aliasing error (zero padding for half of the modes). These operations provide longer running time and larger memory usage in the case with shear. A good balance between lowering time/memory consumption and covering all the physically significant modes is obtained using $\mathrm{56\times 56\times 24}$ modes, such that even in those cases with shear one reaches physical length scales down to just 3 times the Larmor radius (for $l,m=\pm 14$, $l$ and $m$ being radial and poloidal mode numbers).  

\section{Results}\label{sec3}

The simulations results for the three magnetic configurations described above are presented in the following subsections. The resulting turbulence features are discussed by outlining the role of the linear and the nonlinear contributions. In particular, we will derive the spectra of the energies $E_\mathrm{\phi}$, $E_{\tilde{p}}$ defined as 
\begin{equation}
E_\phi(\vec{k})=\frac{1}{2}\,|\nabla_\bot\phi|^2(\vec{k})\qquad E_{\tilde{p}}(\vec{k})=\frac{1}{2}|\tilde{p}(\vec{k})|^2
\end{equation}
 and of the corresponding fluxes
\begin{equation}
\left\{\begin{array}{c} 
\partial_t E_\phi = T_\phi + \Gamma_{\mathrm{drift},\phi}  - \Gamma_{\mathrm{visc}}\hspace{.5cm}\\
\partial_t E_{\tilde{p}} = T_{\tilde{p}} + \Gamma_{\hat{k}} + \Gamma_{\mathrm{drift},\tilde{p}} - \Gamma_{\mathrm{diff}}   
\end{array}
\right.\label{EES}
\end{equation}
where the two drift contributions have no definite sign\footnote{A generalized energy $E_\phi + (B/D)\,E_{\tilde{p}}$ can be defined for which the overall drift contribution is dissipative. We will not introduce it since we want to keep separated the contributions for $\phi$ and $\tilde{p}$ as they correspond to different physical quantities in tokamak experiments with dedicated measurement apparatus.} and read
\begin{equation}
\begin{array}{c} 
\Gamma_{\mathrm{drift},\phi}(\vec{k})=-\displaystyle\frac{BF}{\eta}\, \phi(-\vec{k})\nabla_\parallel^2(\tilde{p}-\phi)(\vec{k}) \\
\Gamma_{\mathrm{drift},\tilde{p}}(\vec{k})=\displaystyle\frac{DF}{\eta}\,\tilde{p}(-\vec{k})\nabla_\parallel^2(\tilde{p}-\phi)(\vec{k})\,,
\end{array}
\end{equation}
while the flux due to the background-pressure gradient is given by the expression
\begin{equation}
\Gamma_{\hat{k}}(\vec{k})=-\hat{k}\,\tilde{p}(-\vec{k})\delta_{y}\phi(\vec{k})\,,
\end{equation}
and the two dissipative terms have the minus signs corresponding to the viscous and the diffusive ones, {\it i.e.}
\begin{equation}
\Gamma_{\mathrm{visc}}=\nu\, |\Delta_\bot\phi(\vec{k})|^2\qquad
\Gamma_{\mathrm{diff}}= \chi_\bot\, |\nabla_\bot\tilde{p}(\vec{k})|^2\,.  
\end{equation}
All the derivative terms above (namely $\nabla_\bot\phi$, $\nabla_\parallel(\tilde{p}-\phi)$, $\delta_{y}\phi$, ..) can be written in FS as simple multiplicative operators ($\mathrm{i}k_\bot\phi(\vec{k})$, $-k_\parallel(\tilde{p}-\phi)(\vec{k})$, $\mathrm{i}k_y\phi(\vec{k})$, ..) only in the two cases without shear, while in the case with shear they are computed going back and forth to real space as described at the end of the previous section, mimicking the procedure adopted for the nonlinear transfer functions $T_\phi$ and $T_{\tilde{p}}$, whose expressions read
\begin{equation}
T_{\phi}(\vec{k})= \phi(-\vec{k})\{\phi,\Delta_\bot \phi\}(\vec{k})\qquad
T_{\tilde{p}}(\vec{k})= \tilde{p}(-\vec{k})\{\phi,\tilde{p}\}(\vec{k})\,.  
\label{Ts}
\end{equation} 
In what follows we will discuss the toroidal cascade to axisymmetric $n=0$ modes, corresponding to $k_z=0$. In this respect, we will split the nonlinear contribution into the 2D part $T^{(2D)}_{\mathrm{..}}$, in which all the fields in Eqs.(\ref{Ts}) are computed at $n=0$, and the remaining part $T^{(n\neq0\rightarrow n=0)}_{\mathrm{..}}$ giving the amount of transition from $n\neq0$ to $n=0$, 
\begin{equation}
T_{\mathrm{..}}(k_x,k_y,0)=T^{(2D)}_{\mathrm{..}}(k_x,k_y)+T^{(n\neq0\rightarrow n=0)}_{\mathrm{..}}(k_x,k_y)\,.    
\end{equation}

The basic free-energy flow goes as follows. $\Gamma_{\hat{k}}$ provides a free-energy source for $E_{\tilde{p}}$ whose magnitude is proportional to $\hat{k}$ and is modulated by $\delta_{y}\phi$, then such energy redistributes by toroidal and poloidal cascades while being dissipated by diffusion. The drift term provides at the same time dissipation and a two-way flow between $E_{\tilde{p}}$ and $E_\phi$ with the latter being redistributed by poloidal and toroidal cascades and dissipated by viscosity.

\subsection{SLAB}\label{sec3.1}

The SLAB case is characterized by fully toroidal magnetic field with $\hat{b}_x=\hat{b}_y=0$ and $\hat{b}_z=1$. The perpendicular Laplacian operator in FS is $\Delta_\bot=-(k_x^2+k_y^2) $, the drift term vanishes for axisymmetric $n=0$ modes and the free-energy source $\Gamma_{\hat{k}}$ is proportional to the poloidal number $m$. 

The simulation results are shown in the multi-panel figure Tab.\ref{frames_SLAB}, where in each panel the values of $\tilde{p}$ and $\phi$ are shown at a given time $t=0.167\mathrm{ms}$ in the poloidal plane $(x,y)$ (figures above) and in the radial-toroidal plane $(x,z)$ (figures below). The two upper/lower panels are obtained for strong/weak background pressure gradient $\ell_0=2\mathrm{cm}$/$\ell_0=10\mathrm{cm}$ with neoclassical (left) and classical (right) diffusivity coefficients $\chi_\bot$.  

\begin{table}
\begin{tabular}{|c|c|}

      \hline \\
            \includegraphics[trim={3.5cm 1.5cm 3.5cm 1.7cm},clip,width=80mm]{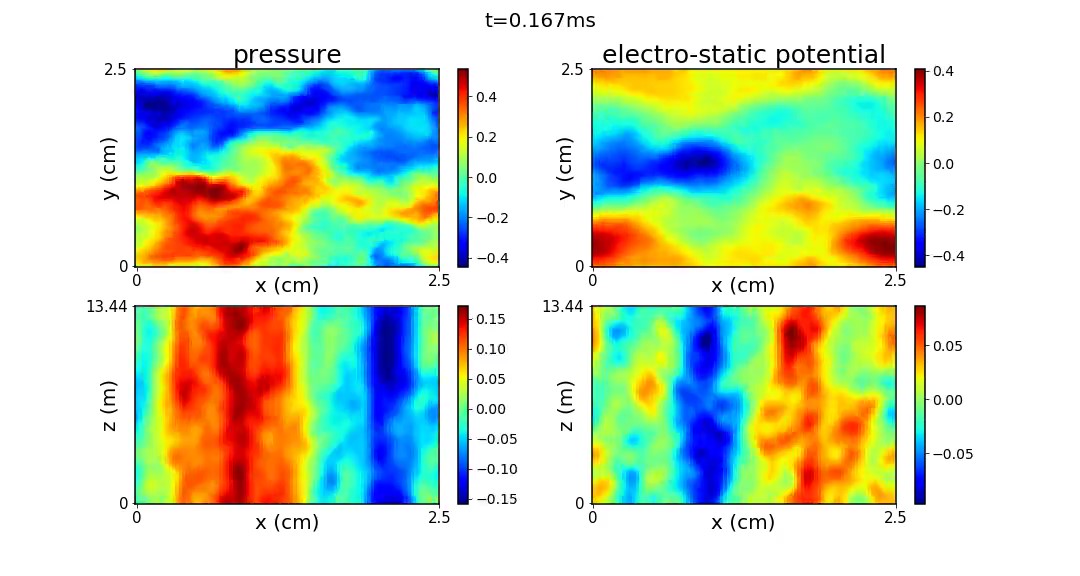} & \includegraphics[trim={3.5cm 1.5cm 3.5cm 1.7cm},clip,width=80mm]{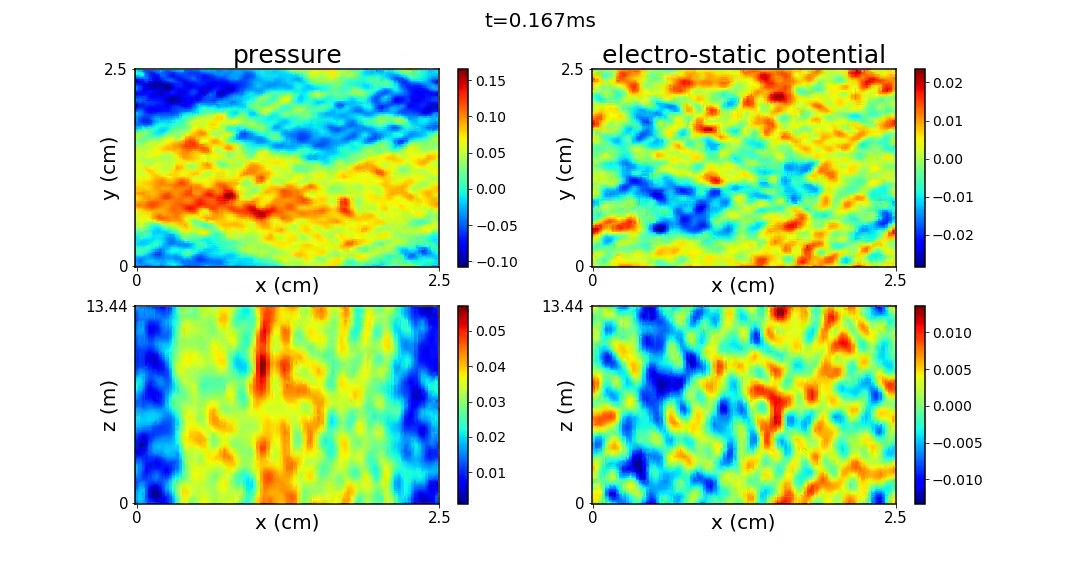} \\
			{\color{red}\textbullet} a) $\chi_{\mathrm{neocl}}$ $\ell_0$=2cm  & {\color{ForestGreen}\textbullet} b) $\chi_{\mathrm{class}}$ $\ell_0$=2cm \\
			\hline \\
      \includegraphics[trim={3.5cm 1.5cm 3.5cm 1.7cm},clip,width=80mm]{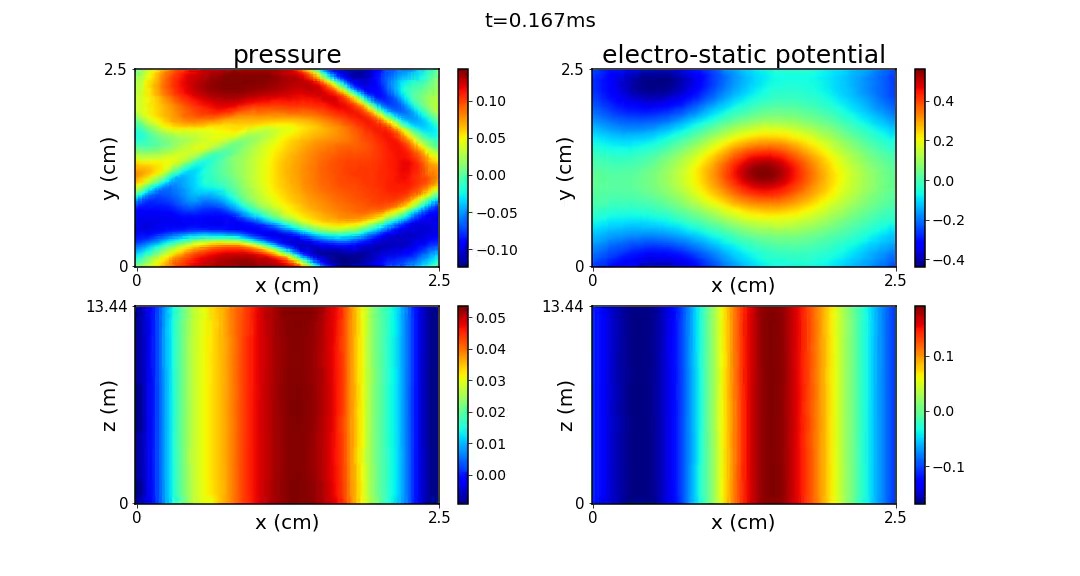} & \includegraphics[trim={3.5cm 1.5cm 3.5cm 1.7cm},clip,width=80mm]{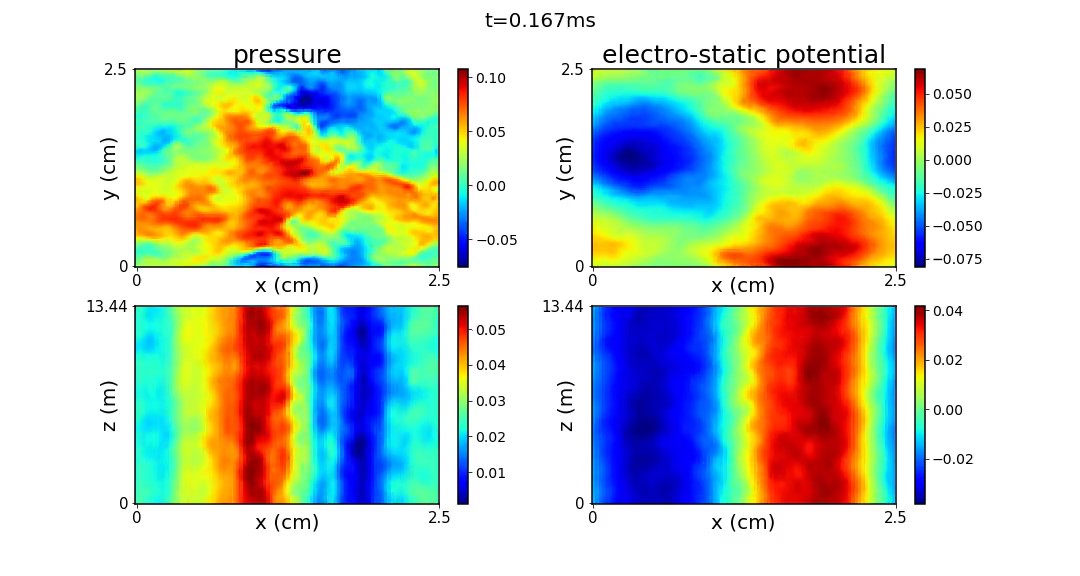} \\
			{\color{blue}\textbullet} c) $\chi_{\mathrm{neocl}}$ $\ell_0$=10cm & {\color{Goldenrod}\textbullet} d) $\chi_{\mathrm{class}}$ $\ell_0$=10cm \\			
			\hline 
\end{tabular}
\caption{The simulation results at time $t=0.167ms$ in SLAB for fully periodic BC. In each panel the figures above contains the poloidal sections, $(x,y)$ plane, of normalized $\tilde{p}$ (left) and $\phi$ (right) while the figures below are for the radial-toroidal $(x,z)$ plane. The corresponding colorbars are shown next to each plot.}
\label{frames_SLAB}
\end{table}

It is worth noting how both the perturbed pressure and electro-static potential reach an axisymmetric configuration in case c) (left bottom panel in Tab.\ref{frames_SLAB}) with the formation of large vortex structures in the poloidal plane. Increasing background pressure gradient (bottom to top) or decreasing diffusivity (left to right) provides a large degree of toroidal asymmetry and the formation of small scale structures in the poloidal plane. These qualitative features are common to all the considered magnetic geometries (see below). 

\begin{figure}[h]
\centering
\includegraphics[width=\textwidth]{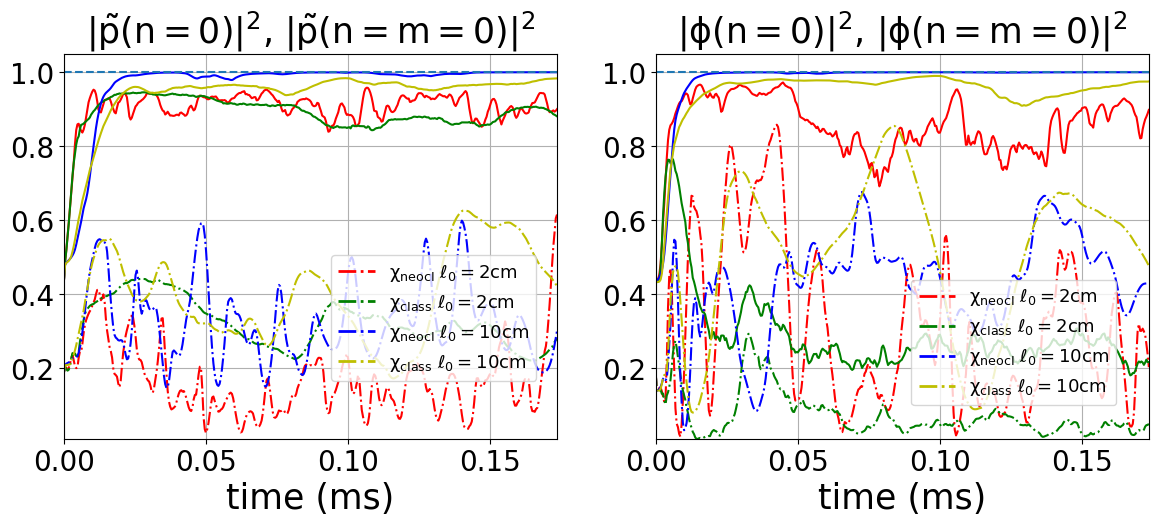}
\caption{The ratio of $n=0$ (solid lines) and of $n=m=0$ (dash-dotted lines) mode amplitudes over the total amplitude for $\tilde{p}$ (left) and $\phi$ (right) vs time (in ms) for the four considered cases in SLAB with fully periodic BC.}
\label{modes_SLAB}
\end{figure}

The fraction of toroidally symmetric mode is shown in figure \ref{modes_SLAB}, where the ratios of $n=0$ (solid lines) and $n=m=0$ (dash-dotted lines) to total mode amplitude are plotted in time for the four considered cases. The former reaches saturated values for both $\phi$ and $\tilde{p}$ that are practically one in case c) (blue) and very close to one in case d) (yellow), outlining how increasing diffusivity reduces the fraction of toroidally asymmetric modes. This is even more evident for $\phi$ amplitude in those cases (a) and b)) with strong background pressure gradients. In fact, in the neoclassical case a) (red) toroidally asymmetric modes are still strongly suppressed with more than $\sim$80\% of $\phi$ amplitudes at $n=0$, while the same ratio reduces to to just $\sim$20\% in case b) (green). The fraction of $n=0$ modes that are also poloidally symmetric ($m=0$) are plotted with dash-dotted lines and their saturated values oscillate around $1/3$ and one half of total $n=0$ mode amplitude for $\tilde{p}$ and $\phi$, respectively. 

\begin{figure}[h]
\centering
\includegraphics[width=\textwidth]{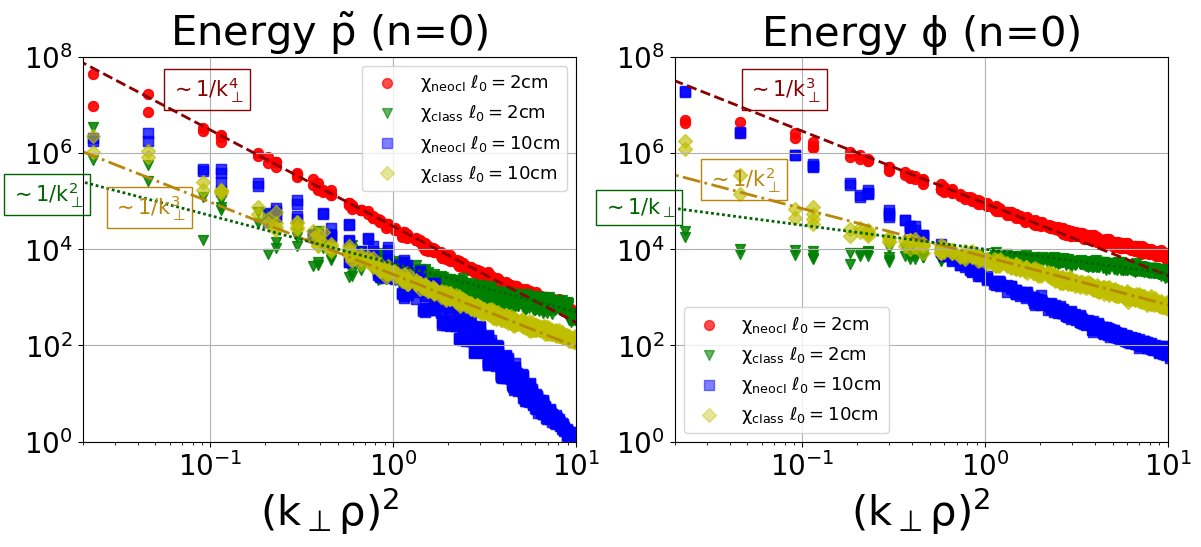}
\caption{The scatter plots of normalized axisymmetric spectral energies $E_{\tilde{p}}$ (left) and $E_\phi$ (right) vs $(k_\bot\rho)^2$ for the four considered cases in SLAB with fully periodic BC. They are computed by averaging over a time window of $\sim 0.13ms$. Some curves are also plotted that approximate quite well some of the spectra profiles and which behave as: $1/k_\bot^4$ (dashed dark red line), $1/k_\bot^3$ (dash-dotted dark yellow line) and $1/k_\bot^2$ (dotted dark green line) for $E_{\tilde{p}}$ (left) and $1/k_\bot^3$ (dashed dark red line), $1/k_\bot^2$ (dash-dotted dark yellow line) and $1/k_\bot$ (dotted dark green line) for $E_\phi$ (right).}
\label{en_spectra_SLAB}
\end{figure}

The energy spectra at $n=0$ for $\tilde{p}$ and $\phi$ are presented in Fig.\ref{en_spectra_SLAB}. In order to see potential anisotropies, it is shown the scatter plot for all the $(k_\bot\rho)^2$ values obtained from the computational discrete Cartesian grid in FS (anisotropies would appear as different spectral values for the same $(k_\bot\rho)^2$). The 2D spectra are indeed quite isotropic and show a strong dependence on the chosen parameters. In particular, for short wave-lengths ($k_\bot\rho>1$) $E_\phi$ goes as $\sim 1/k_\bot$ in case b) (green), $\sim 1/k_\bot^2$ in case d) (yellow) and $1/k_\bot^3$ in case a) (red) and the corresponding $E_{\tilde{p}}$ spectra are suppressed by a factor $1/k_\bot$. The spectra for case c) (blue) decays more rapidly with $k_\bot\rho$ and differently from the others, that are quite stationary, it becomes even more peaked moving forward the time window.

\begin{figure}[h]
\centering
\includegraphics[width=\textwidth]{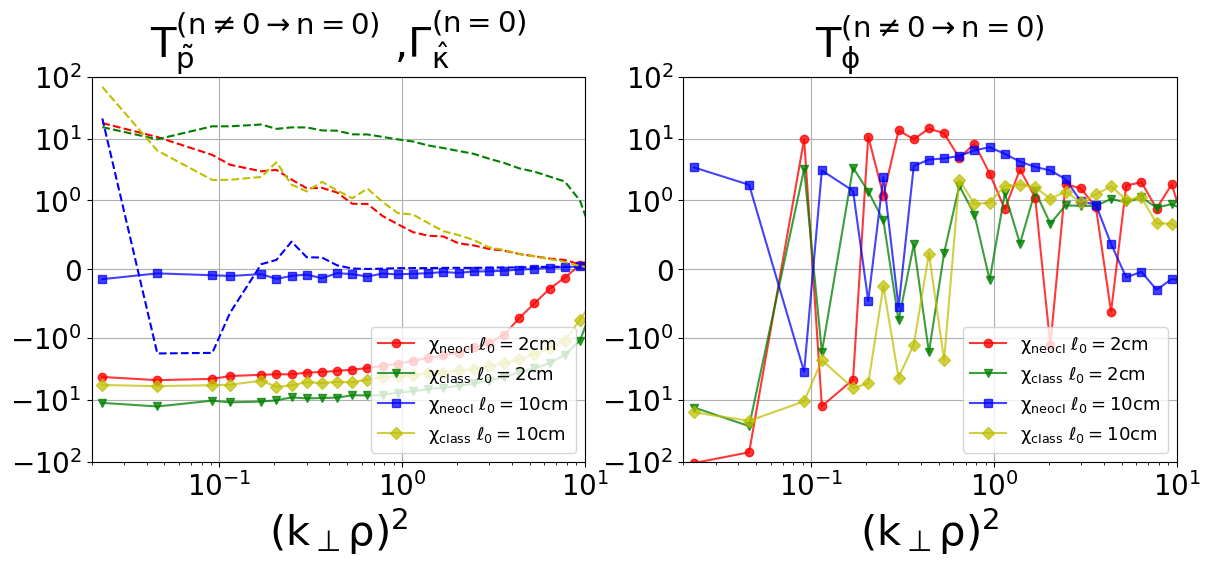}
\caption{The nonlinear transfer functions of toroidal transitions $n\neq0\rightarrow n=0$ (solid lines) for $\tilde{p}$ (left) and $\phi$ (right) vs $(k_\bot\rho)^2$ and the flux due to background pressure gradient (dashed lines, left) for the four considered cases in SLAB with fully periodic BC. Each plotted value is normalized with the corresponding dissipative flux, {\it i.e.} the diffusivity and the viscous contributions for $\tilde{p}$ and $\phi$, respectively. The chosen values of $(k_\bot\rho)^2$ are constructed from the computational Cartesian grid in FS by taking the five smallest values of $k_\bot$ and by logarithmically sampling the relic part.}
\label{Tpphi_SLAB}
\end{figure}

In order to understand the impact of 3D toroidal transitions on 2D spectra, we plot in Fig.\ref{Tpphi_SLAB} the nonlinear transfer functions $T^{(n\neq0\rightarrow n=0)}_{\tilde{p}}$ (left) and $T^{(n\neq0\rightarrow n=0)}_{\phi}$ (right) accounting for the transitions from $n\neq 0$ to $n=0$. We adopt a logarithmic sampling for $(k_\bot\rho)^2$ for readability. Each transfer function is normalized with the corresponding dissipative contribution ($\Gamma_{\mathrm{diff}}$ for $\tilde{p}$ and $\Gamma_{\mathrm{visc}}$ for $\phi$). 

\begin{figure}[h]
\centering
\includegraphics[width=\textwidth]{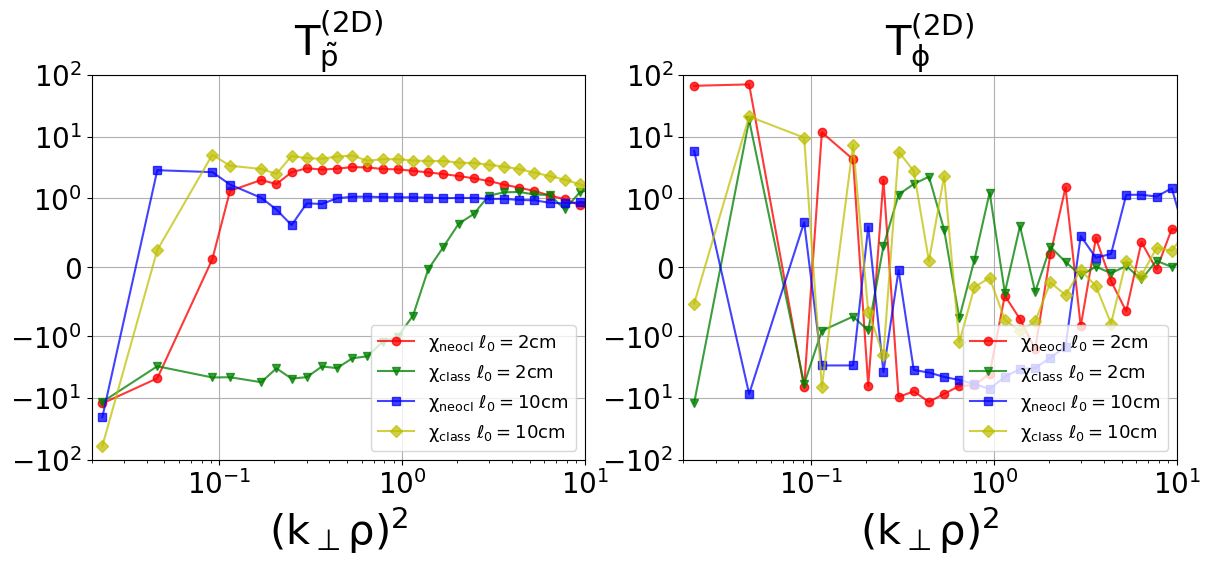}
\caption{The nonlinear transfer functions of 2D transitions $n=0\rightarrow n=0$ (solid lines) for $\tilde{p}$ (left) and $\phi$ (right) vs $(k_\bot\rho)^2$ for the four considered cases in SLAB with fully periodic BC. Each plotted value is normalized with the corresponding dissipative flux, {\it i.e.} the diffusivity and the viscous contributions for $\tilde{p}$ and $\phi$, respectively.}
\label{T2D_SLAB}
\end{figure}

It is worth noting that $T^{(n\neq0\rightarrow n=0)}_{\tilde{p}}$ is negative, thus $n\neq 0$ modes behave as a sink (direct toroidal cascade) for pressure that is more intense by increasing background pressure gradients (green and red vs yellow and blue) and reducing diffusivity (blue and red vs yellow and green). In particular, for large part of the $k_\bot$ spectrum $T^{(n\neq0\rightarrow n=0)}_{\tilde{p}}<-1$ meaning that it is larger than diffusion thus it is the main dissipative contribution partially compensating the source due to background pressure gradients $\Gamma_{\hat{k}}$ shown in dashed lines in Fig.\ref{Tpphi_SLAB} (left). The corresponding 2D transition functions are shown in fig.\ref{T2D_SLAB} (left). They increase from negative to positive values while increasing $k_\bot$, thus they provide an energy flow from large to small wave-lengths. 

On the contrary, $T^{(n\neq0\rightarrow n=0)}_{\phi}$ has no definite sign and it plays the role of source or sink depending on the sampled $k_\bot$ value and on the considered case. We can notice how it is close to 1 for $k_\bot\rho>1$ in case d) (yellow), signaling that there is an inverse toroidal cascade partially compensating the viscous contribution and making the energy spectrum close to a pure 2D Euler inviscid one ($\sim 1/k_\bot^2$, Kraichman-Kolmogorow spectrum \cite{Kraichnan:80}\footnote{A factor $1/k_\bot$ is due to the transformation from Cartesian $k_x, k_y$ to cylindrical coordinates, reproducing the well-known formula $1/k_\bot^3$.} obtained for the reduced drift turbulence model in Ref.\cite{Montani:22}). In the same part of the spectrum the red curve exhibits strong fluctuations around 0 amplitude, thus the overall contribution of the nonlinear toroidal transfers is smaller than one and this could explain the behavior $E_\phi\sim 1/k_\bot^3$ (Saffman spectrum \cite{Saffman:71}, see the footnote). For smaller wave-numbers toroidal and 2D transfer functions (the latter shown in Fig.\ref{T2D_SLAB}) are strongly anticorrelated with large fluctuations between neighbor sampled $k_\bot$ values. This means that 3D transitions are as relevant as 2D ones in determining the 2D energy spectra in Fig.\ref{en_spectra_SLAB}.   

  
In case c) (blue) one notices that the direct toroidal cascade for $\tilde{p}$ is very low, while there is an inverse toroidal cascade for most of $\phi$ spectra, leading to the axisymmetric profiles in table \ref{frames_SLAB} and strongly peaked spectra in Fig.\ref{en_spectra_SLAB}. In other words, the free-energy source is low (weak $p_0$ gradients) and dissipation is high (neoclassical diffusivity), such that the turbulent source $\Gamma_{\hat{k}}$ is not able to sustain the toroidal cascades and thus drift turbulence decays. In fact, this case is far from stationary, with axisymmetric modes growing at the expenses of asymmetric ones and the 2D spectra getting more and more stiff (condensation). 

\begin{figure}[h]
\centering
\includegraphics[width=.7\textwidth]{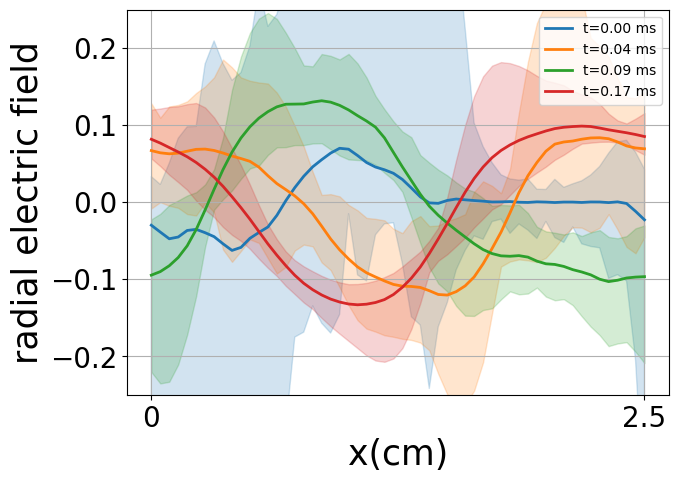}
\caption{The profile of the radial electric field in case c) at four given time values: start (blue), $t=0.04ms$ (orange), $t=0.09ms$ (green) and $t=0.17ms$ (red). The solid lines denote the averaged value along $y$ and $z$ directions, while the corresponding shaded regions cover the whole range of variability along such directions at the given time values.}
\label{rad_elf_SLAB}
\end{figure}

This is accompanied by the formation of a radially-sheared electric field, that seeds further turbulence quench, as shown in Fig.\ref{rad_elf_SLAB} where the toroidally and poloidally averaged radial electric field (solid lines) and the corresponding range of variability around the average (shaded regions) is plot along the $x$ radial direction at four time values. An averaged sheared profile arises from the start (blue) to the end (red) of the simulation with the shaded region getting narrow with time signaling decreasing variations from average along poloidal and toroidal directions.  

\subsubsection{Dirichlet BC}

The results for the simulations in SLAB with vanishing radial Dirichlet BC implemented through the method discussed in section \ref{sec2} are presented in the multi-panel figure Tab.\ref{frames_SLAB_NEW}. The physical solution corresponds to the profiles in the first half of x-axis and the second half just contains a specular copy.    

\begin{table}
\begin{tabular}{|c|c|}

      \hline \\
            \includegraphics[trim={3.5cm 1.5cm 3.5cm 1.7cm},clip,width=80mm]{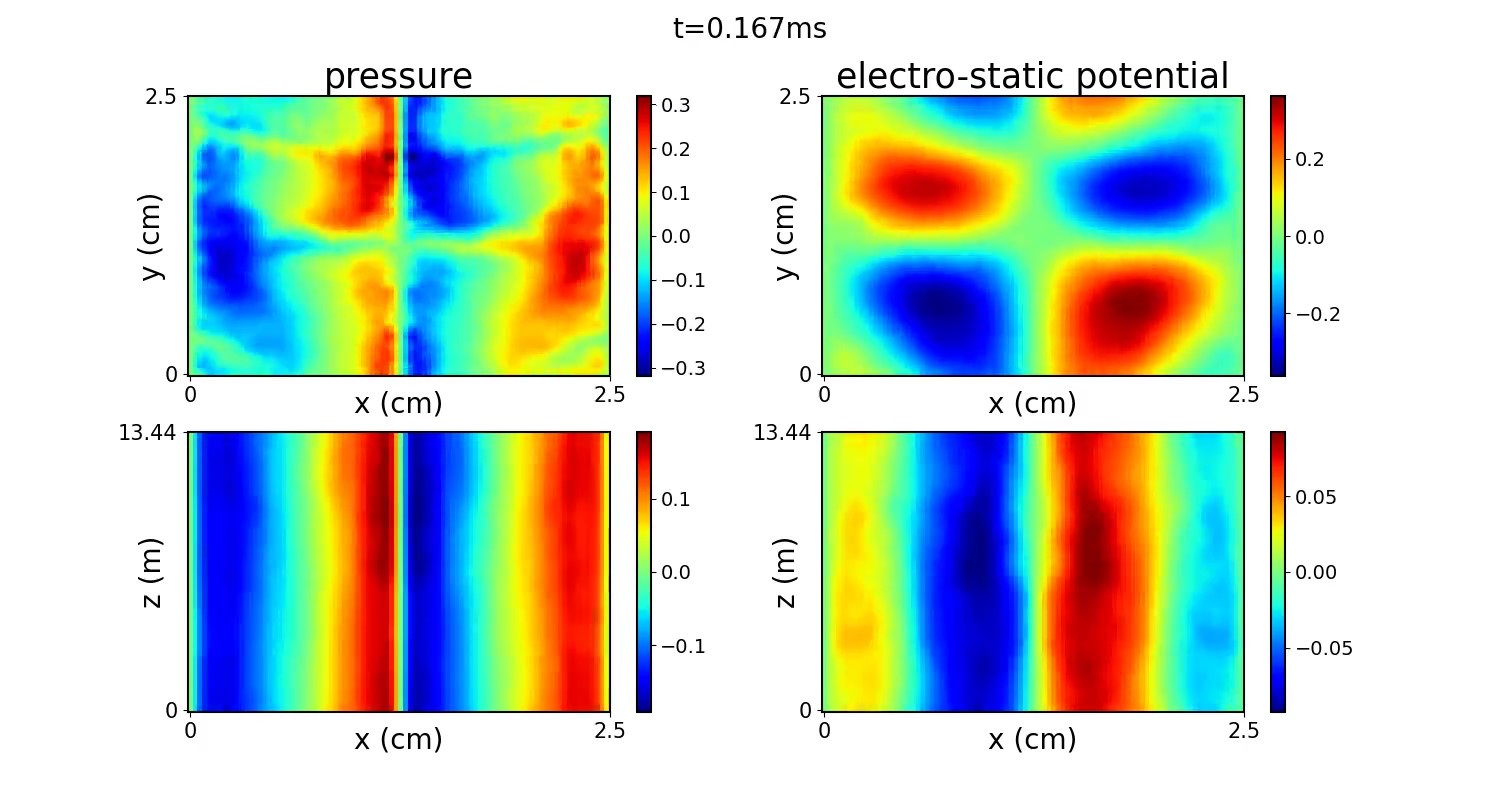} & \includegraphics[trim={3.5cm 1.5cm 3.5cm 1.7cm},clip,width=80mm]{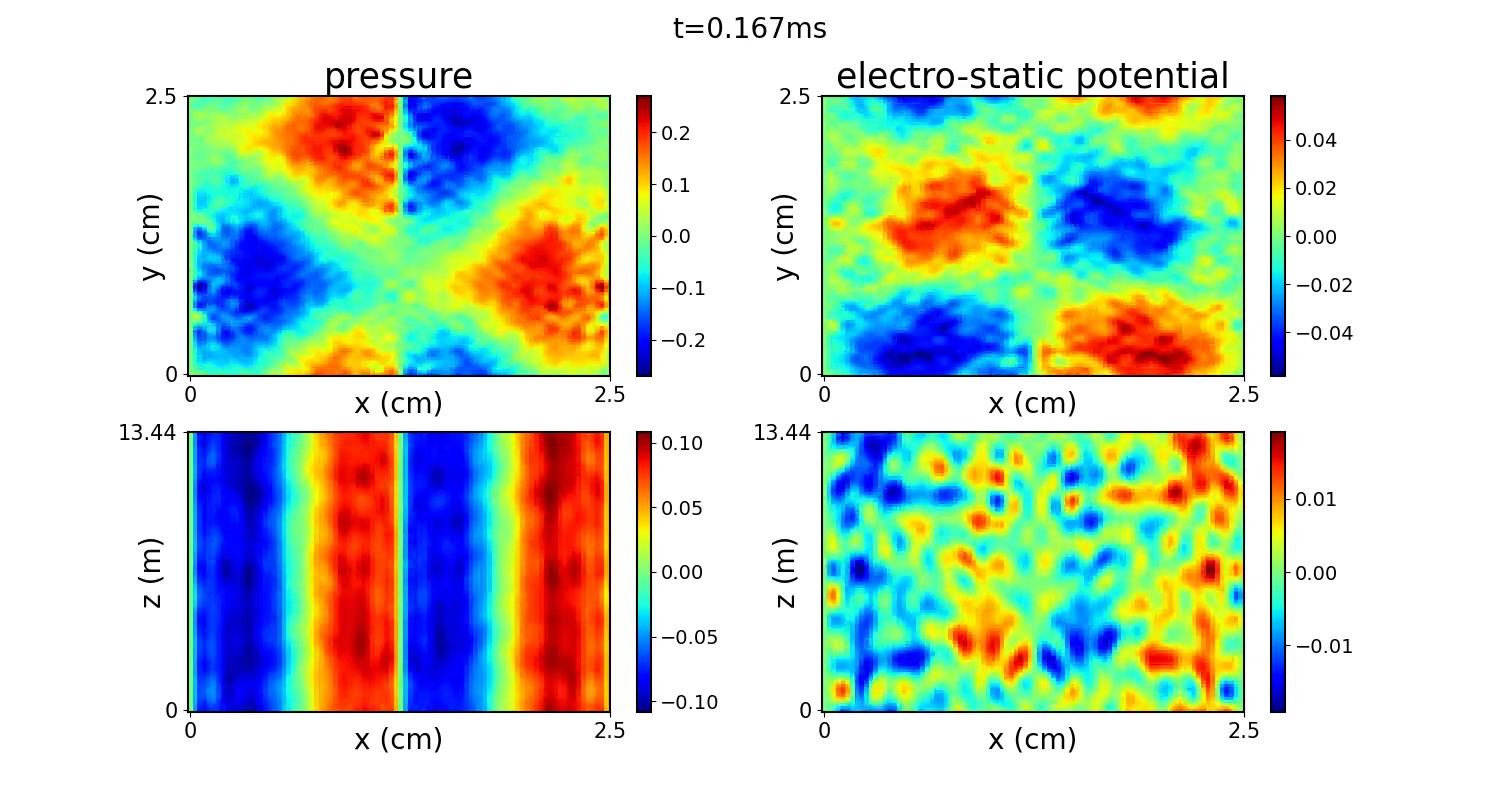} \\
			{\color{red}\textbullet} a) $\chi_{\mathrm{neocl}}$ $\ell_0$=2cm  & {\color{ForestGreen}\textbullet} b) $\chi_{\mathrm{class}}$ $\ell_0$=2cm \\
			\hline \\
      \includegraphics[trim={3.5cm 1.5cm 3.5cm 1.7cm},clip,width=80mm]{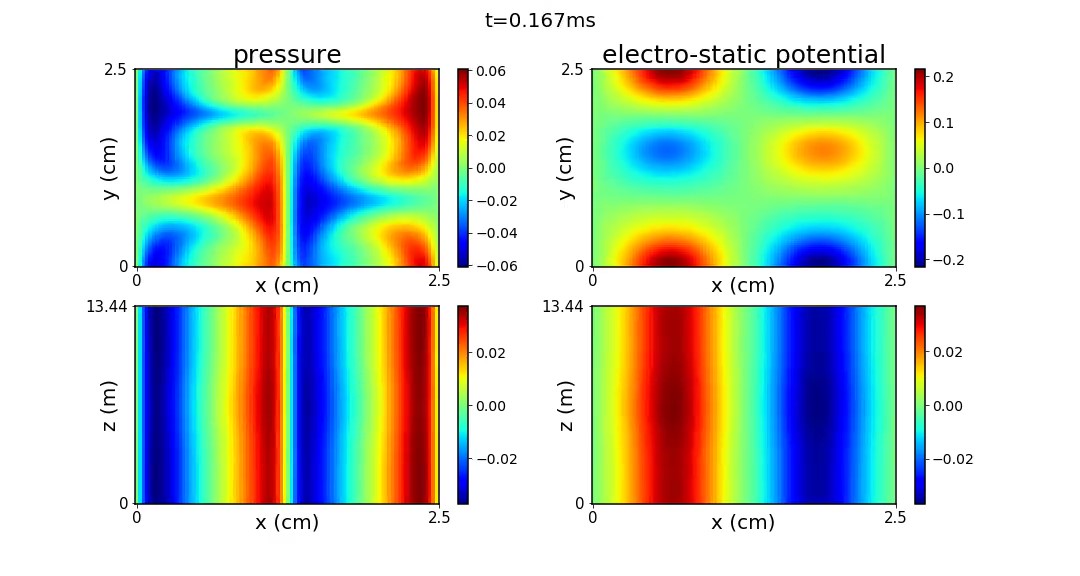} & \includegraphics[trim={3.5cm 1.5cm 3.5cm 1.7cm},clip,width=80mm]{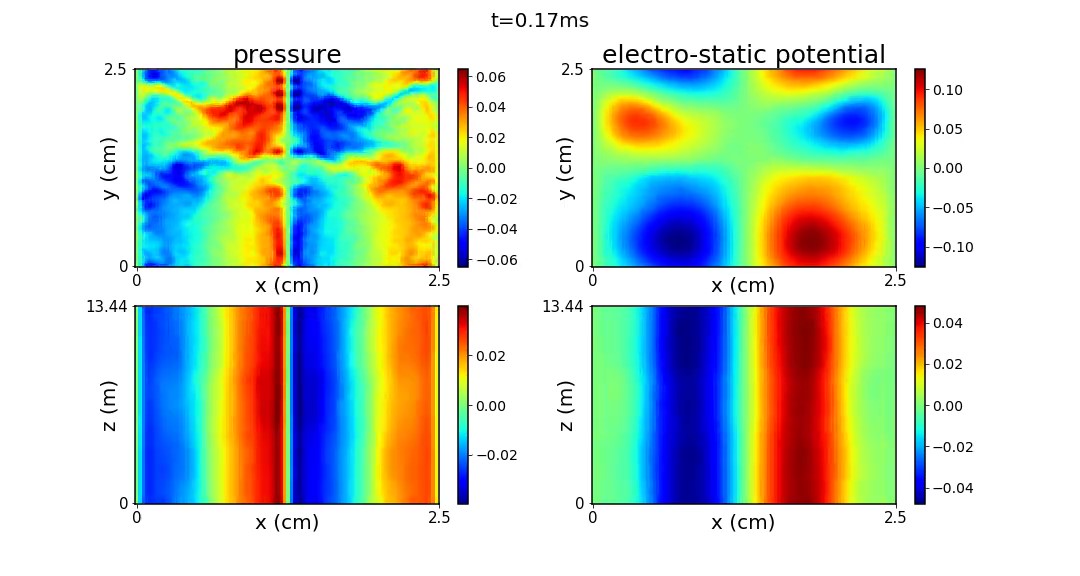} \\
			{\color{blue}\textbullet} c) $\chi_{\mathrm{neocl}}$ $\ell_0$=10cm & {\color{Goldenrod}\textbullet} d) $\chi_{\mathrm{class}}$ $\ell_0$=10cm \\			
			\hline 
\end{tabular}
\caption{The simulation results at time $t=0.167ms$ in SLAB for vanishing radial Dirichlet BC. In each panel the figures above contain the poloidal sections, $(x,y)$ plane, of normalized $\tilde{p}$ (left) and $\phi$ (right) while the figures below are for the radial-toroidal $(x,z)$ plane. The corresponding colorbars are shown next to each plot.}
\label{frames_SLAB_NEW}
\end{table}

The degree of toroidal symmetry outlines the same qualitative behavior as in the case with fully periodic BC: the most symmetric configuration is obtained in case c) (left bottom panel), with increasing asymmetries while increasing background pressure gradients (bottom to top) and/or reducing diffusivity (left to right). One can also notice the formation of elongated structures in $y$ direction mostly for $\tilde{p}$.
 
\begin{figure}[h]
\centering
\includegraphics[width=\textwidth]{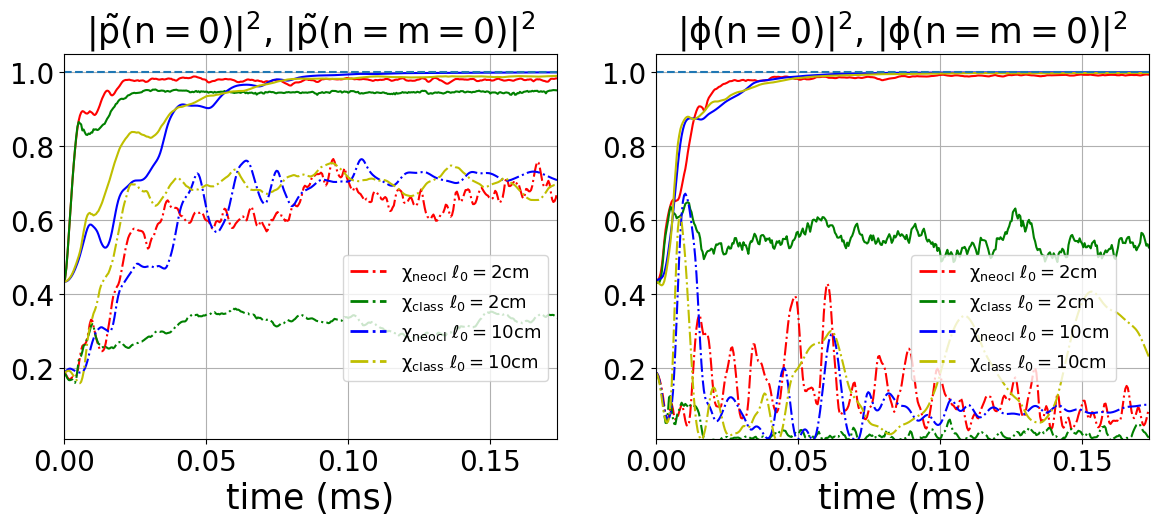}
\caption{The ratio of $n=0$ (solid lines) and of $n=m=0$ (dash-dotted lines) mode amplitudes over the total amplitude for $\tilde{p}$ (left) and $\phi$ (right) vs time (in ms) for the four considered cases in SLAB with vanishing radial Dirichlet BC.}\label{modes_SLAB_NEW}
\end{figure}

The ratio of $n=0$ (solid lines) and $n=m=0$ (dash-dotted lines) mode amplitudes to the total amplitude is plotted in time in fig.\ref{modes_SLAB_NEW}. The solid lines for $\tilde{p}$ (left) are all close to one, signaling that $n\neq 0$ modes are generically suppressed and except in case b) (green) the dashed lines are quite high, more than 50\%, meaning that the saturated modes are mostly poloidally symmetric too. For $\phi$ (right) one still get strong degree of toroidal symmetry in all cases but b) (green) while the degree of poloidal symmetry is much lower than that for $\tilde{p}$. 

\begin{figure}[h]
\centering
\includegraphics[width=\textwidth]{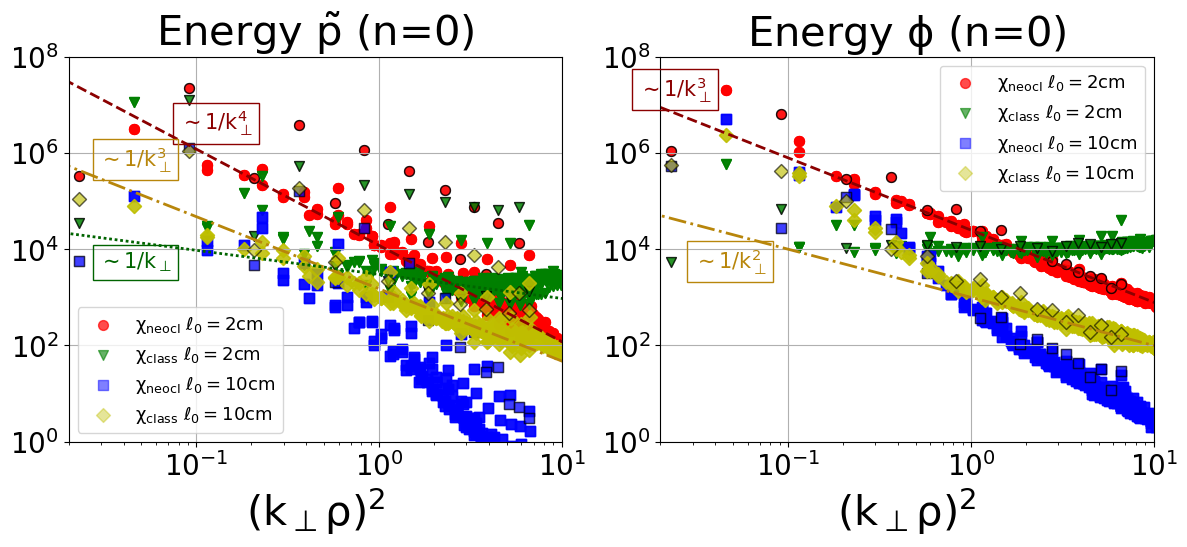}
\caption{The scatter plots of normalized axisymmetric spectral energies $E_{\tilde{p}}$ (left) and $E_\phi$ (right) vs $(k_{\bot}\rho)^2$ for the four considered cases in SLAB with vanishing radial Dirichlet BC. They are computed by averaging over a time window of $\sim 0.13ms$. Black edges are for those points corresponding to $m=0$ modes ($k_y=0$). Some curves are also plotted that approximate quite well some of the spectra profiles and which behave as: $1/k_\bot^4$ (dashed dark red line), $1/k_\bot^3$ (dash-dotted dark yellow line) and $1/k_\bot$ (dotted dark green line) for $E_{\tilde{p}}$ (left) and $1/k_\bot^3$ (dashed dark red line) and $1/k_\bot^2$ (dash-dotted dark yellow line) for $E_\phi$ (right).}
\label{en_spectra_NEW_SLAB}
\end{figure}

The emergence of elongated structures in $y$ direction is confirmed by the enhancement of the energy spectra for the poloidally symmetric $m=0$ modes, emphasized graphically by black edges, in Fig.\ref{en_spectra_NEW_SLAB}. Generically, the spectra are mildly anisotropic, especially $E_{\tilde{p}}$ one, and a part from that they are quite similar to those with fully periodic BC (see the dark red and dark yellow lines in Fig.\ref{en_spectra_NEW_SLAB} reproducing the spectra in Fig.\ref{en_spectra_SLAB} for cases a) and d)), except in case b) (green) in which $E_\phi$ is almost constant and $E_{\tilde{p}}$ goes approximately as $1/k_\bot$ (left, dark green line).

\begin{figure}[h]
\centering
\includegraphics[width=\textwidth]{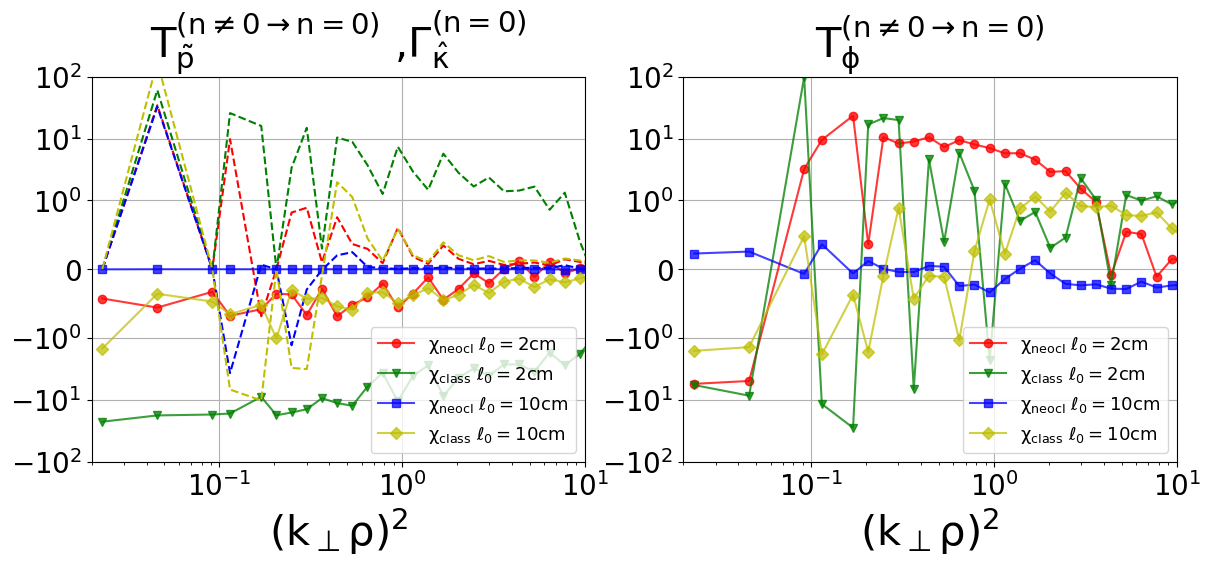}
\caption{The nonlinear transfer functions of toroidal transitions $n\neq0\rightarrow n=0$ (solid lines) for $\tilde{p}$ (left) and $\phi$ (right) vs $(k_\bot\rho)^2$ and the flux due to background pressure gradient (dashed lines, left) for the four considered cases in SLAB with vanishing radial Dirichlet BC. Each plotted value is normalized with the corresponding dissipative flux, {\it i.e.} the diffusivity and the viscous contributions for $\tilde{p}$ and $\phi$, respectively. The chosen values of $(k_\bot\rho)^2$ are constructed from the computational Cartesian grid in FS by taking the five smallest values of $k_\bot$ and by logarithmically sampling the relic part.}
\label{Tpphi_NEW_SLAB}
\end{figure}

The toroidal transfer functions $T^{(n\neq0\rightarrow n=0)}_{\tilde{p}}$ and $T^{(n\neq0\rightarrow n=0)}_{\phi}$ in Fig.\ref{Tpphi_NEW_SLAB} show a similar qualitative behavior as with fully periodic BC: $T^{(n\neq0\rightarrow n=0)}_{\tilde{p}}$ is still negative even if less intense at least for the cases a) (red) and d) (yellow) with respect to the those in fig.\ref{Tpphi_SLAB}, $T^{(n\neq0\rightarrow n=0)}_{\phi}$ is positive/negative for short/long wavelengths, respectively. Moreover, the comparison of the dashed lines (left) with those in fig.\ref{Tpphi_SLAB} shows that the flux due to background pressure gradient is generically lower that with fully periodic BC. 

\begin{figure}[h]
\centering
\includegraphics[width=\textwidth]{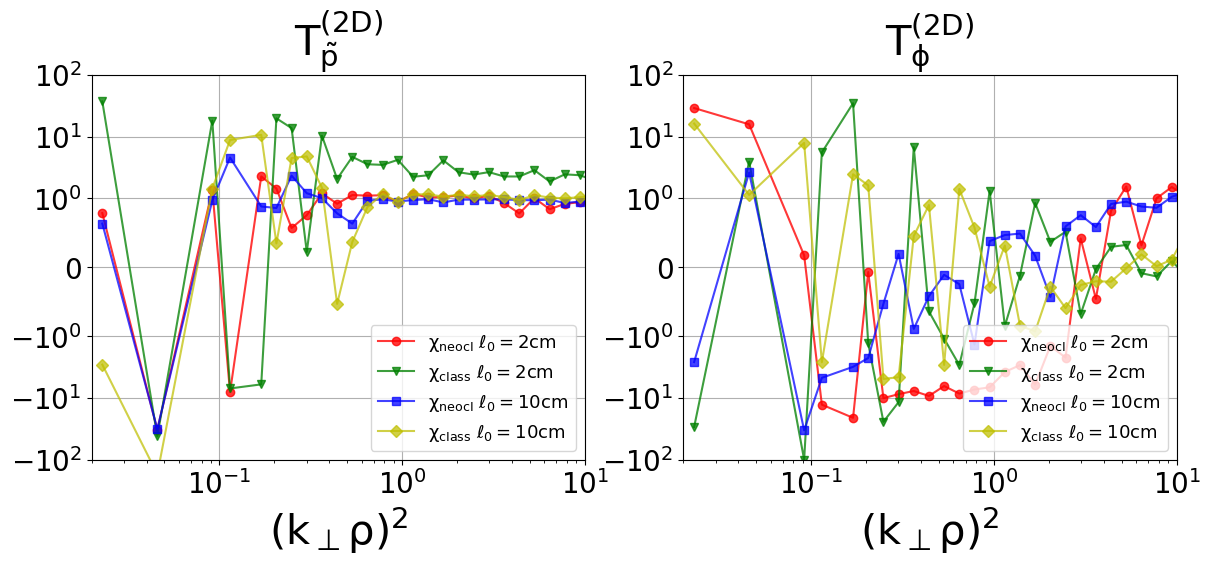}
\caption{The nonlinear transfer functions of 2D transitions $n=0\rightarrow n=0$ (solid lines) for $\tilde{p}$ (left) and $\phi$ (right) vs $(k_\bot\rho)^2$ for the four considered cases in SLAB with vanishing radial Dirichlet BC. Each plotted value is normalized with the corresponding dissipative flux, {\it i.e.} the diffusivity and the viscous contributions for $\tilde{p}$ and $\phi$, respectively.}
\label{T2D_NEW_SLAB}
\end{figure}

2D transfer functions are shown in fig.\ref{T2D_NEW_SLAB}. Their magnitude and that of toroidal transitions are of the same order, while the resulting spectral features are highly non trivial and anticorrelated. 

Case c) is now a real 2D effective model since toroidal transitions are negligible. This can be seen noting that the values of $T^{(n\neq0\rightarrow n=0)}_{..}$ (blue) in Fig.\ref{Tpphi_NEW_SLAB} are close to zero and lower than their 2D counterparts in Fig.\ref{T2D_NEW_SLAB}. 

Therefore, the analysis of the simulations in SLAB outlines how there is an effective inverse toroidal cascade, as toroidally symmetric structures develop. However, the magnitude of such cascade is strongly affected by the value of the dissipative ($\chi_\bot$) and source ($\ell_0$) parameters and non-trivially integrated with 2D transitions leading to various kinds of 2D spectral profiles (that gets wider decreasing diffusivity and increasing background pressure gradients). If vanishing Dirichlet BC are imposed a mild poloidal cascade is also produced leading to the enhancement of poloidally symmetric $m=0$ modes and the reduction of turbulence free-energy source.

\subsection{Constant poloidal field}\label{sec3.2}

We consider now a constant poloidal field, characterized by $B_x=0$ and 
\begin{equation}
\frac{B_y}{B_z}=\alpha\,
\end{equation}
and the constant parameter $\alpha= \frac{L_\bot}{L_\parallel}=1.5\cdot 10^{-3}$ is chosen such that in normalized coordinates the parallel direction is diagonal in $(y,z)$ plane. This choice corresponds to a very large safety factor $q\sim 700$ and it is a reliable model for the SOL region far from the X-point (limiter-like configuration of the tokamak).
The projection of $x$ and $y$ derivatives onto the perpendicular plane are $\delta_x=\partial_x$ and $\delta_y=\partial_y+\hat{b}_y\frac{L_\bot}{L_\parallel}\partial_z\sim \partial_y$ and the perpendicular Laplacian operator in FS reads 
\begin{equation}
\Delta_\bot=k_x^2+\frac{1}{1+\hat{b}_y^2}\left(k_y-\hat{b}_y\frac{L_\bot}{L_\parallel}k_z\right)^2\sim k_x^2+ k_y^2
\label{DeltaperpAC}
\end{equation}
that is close to the expression in SLAB. The main modification is the expression of the parallel derivative that now reads
\begin{equation}
\nabla_\parallel= \frac{1}{\sqrt{1+\hat{b}_y^2}}\left(k_z+ \hat{b}_y\frac{L_\parallel}{L_\bot}k_z\right)
\label{NablaparAC}
\end{equation}
and it depends on $y$ derivatives explicitly breaking isotropy in the poloidal plane. 

The results of the simulations are shown in the multi-panel figure Tab.\ref{frames_AC}. 
\begin{table}
\begin{tabular}{|c|c|}

      \hline \\
            \includegraphics[trim={3.5cm 1.5cm 3.5cm 1.7cm},clip,width=80mm]{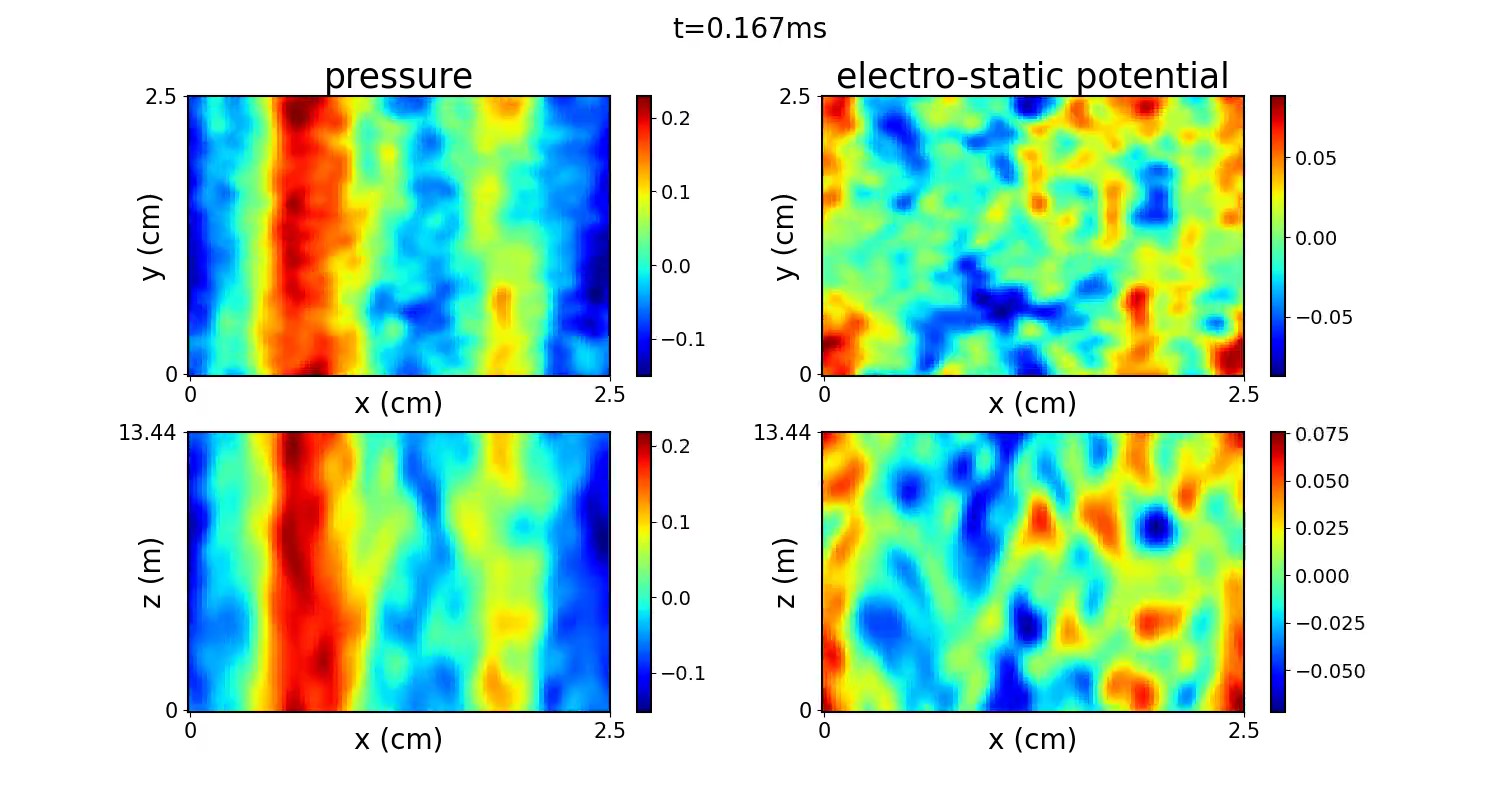} & \includegraphics[trim={3.5cm 1.5cm 3.5cm 1.7cm},clip,width=80mm]{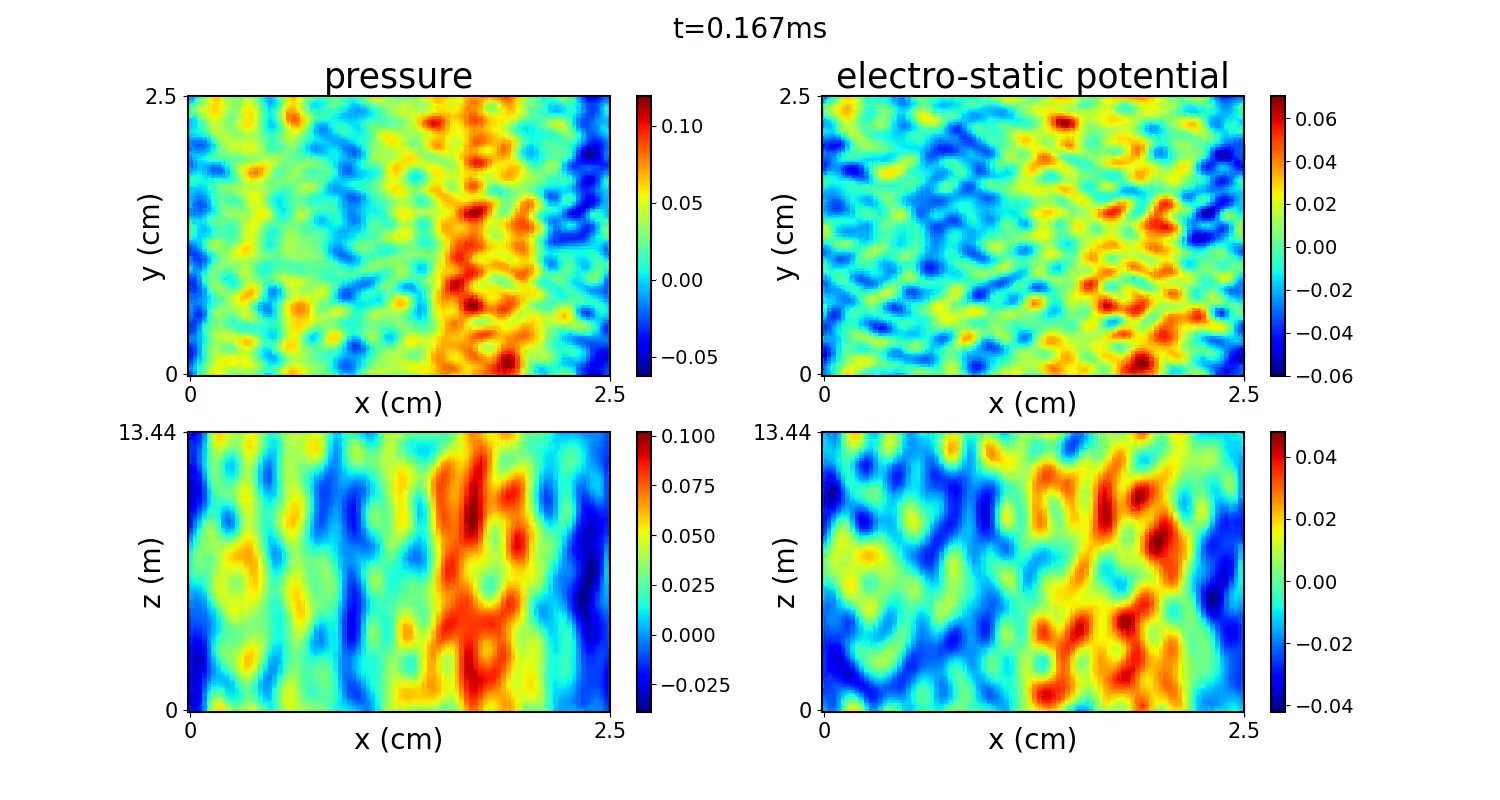} \\
			{\color{red}\textbullet} a) $\chi_{\mathrm{neocl}}$ $\ell_0$=2cm  & {\color{ForestGreen}\textbullet} b) $\chi_{\mathrm{class}}$ $\ell_0$=2cm \\
			\hline \\
      \includegraphics[trim={3.5cm 1.5cm 3.5cm 1.7cm},clip,width=80mm]{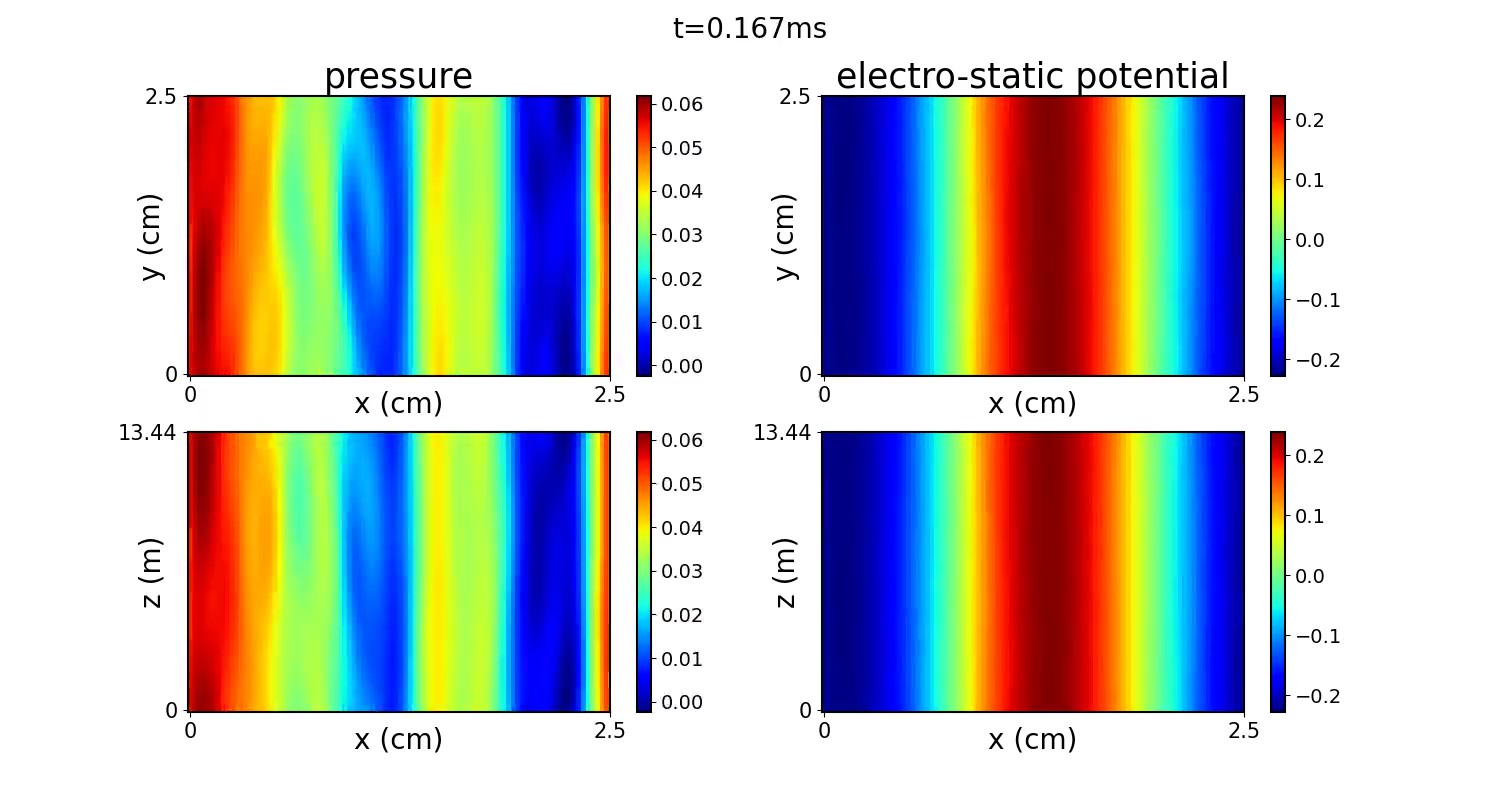} & \includegraphics[trim={3.5cm 1.5cm 3.5cm 1.7cm},clip,width=80mm]{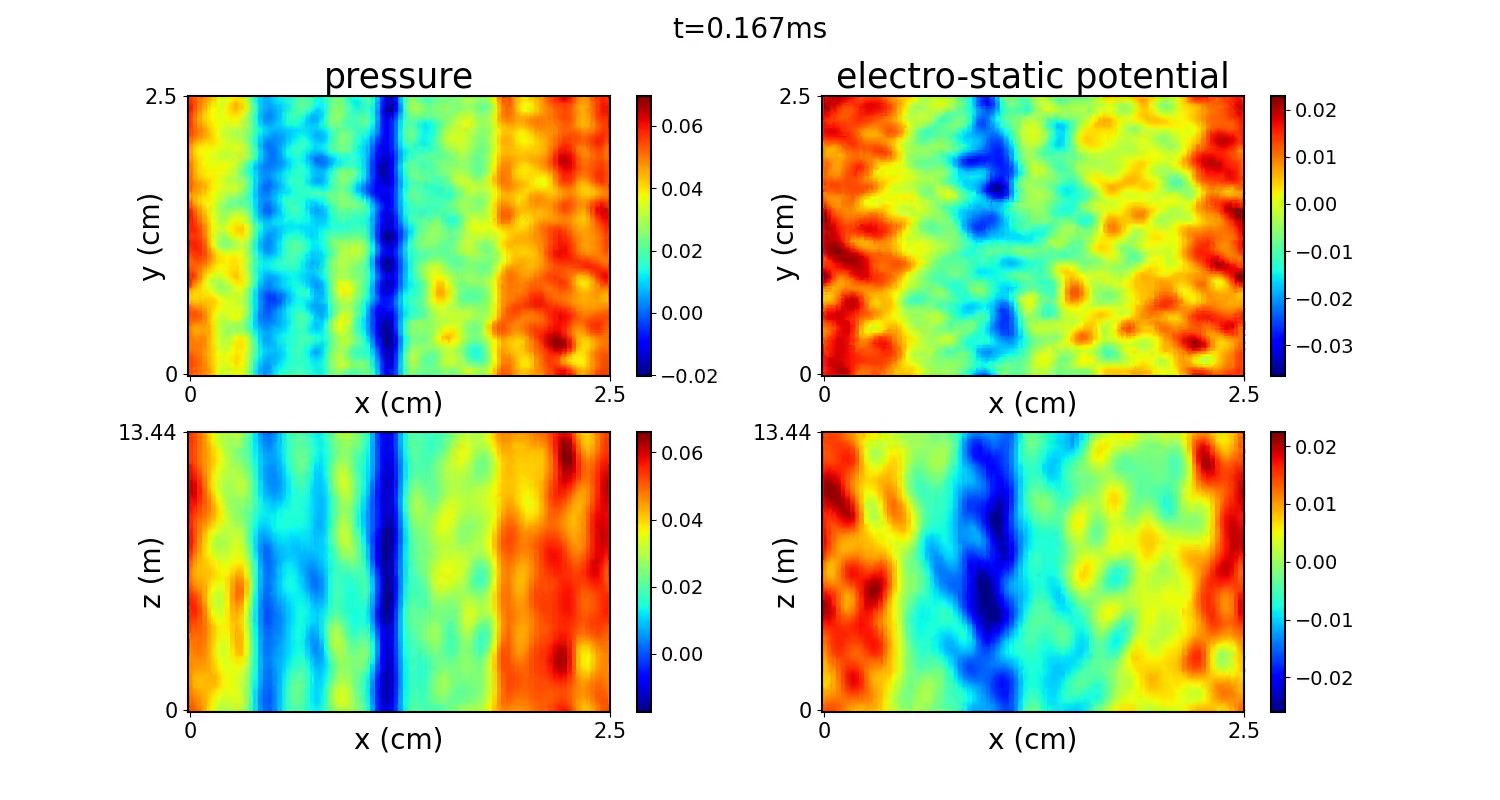} \\
			{\color{blue}\textbullet} c) $\chi_{\mathrm{neocl}}$ $\ell_0$=10cm & {\color{Goldenrod}\textbullet} d) $\chi_{\mathrm{class}}$ $\ell_0$=10cm \\			
			\hline 
\end{tabular}
\caption{The simulation results at time $t=0.167ms$ for constant poloidal magnetic field and fully periodic BC. In each panel the figures above contain the poloidal sections, $(x,y)$ plane, of normalized $\tilde{p}$ (left) and $\phi$ (right) while the figures below are for the radial-toroidal $(x,z)$ plane. The corresponding colorbars are shown next to each plot.}
\label{frames_AC}
\end{table}

One can notice the emergence of configurations with a similar degree of homogeneity along both $z$ and $y$ directions, with cases c) (left bottom panel) and b) (right top panel) being the most and least symmetric ones, respectively. In other words, the inverse toroidal cascade typical of SLAB case here proceeds in conjunction with a similar cascade along the poloidal direction $y$. This is confirmed by:

\begin{figure}[h]
\centering
\includegraphics[width=\textwidth]{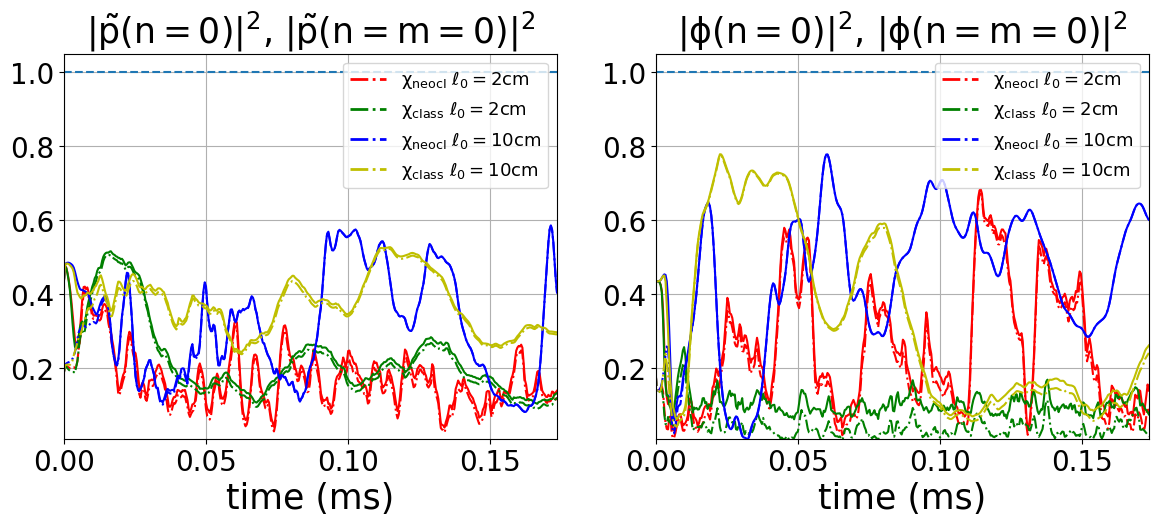}
\caption{The ratio of $n=0$ (solid lines) and of $n=m=0$ (dash-dotted lines) mode amplitudes over the total amplitude for $\tilde{p}$ (left) and $\phi$ (right) vs time (in ms) for the four considered cases with constant poloidal magnetic field and fully periodic BC.}
\label{modes_AC}
\end{figure}
 
\begin{figure}[h]
\centering
\includegraphics[width=\textwidth]{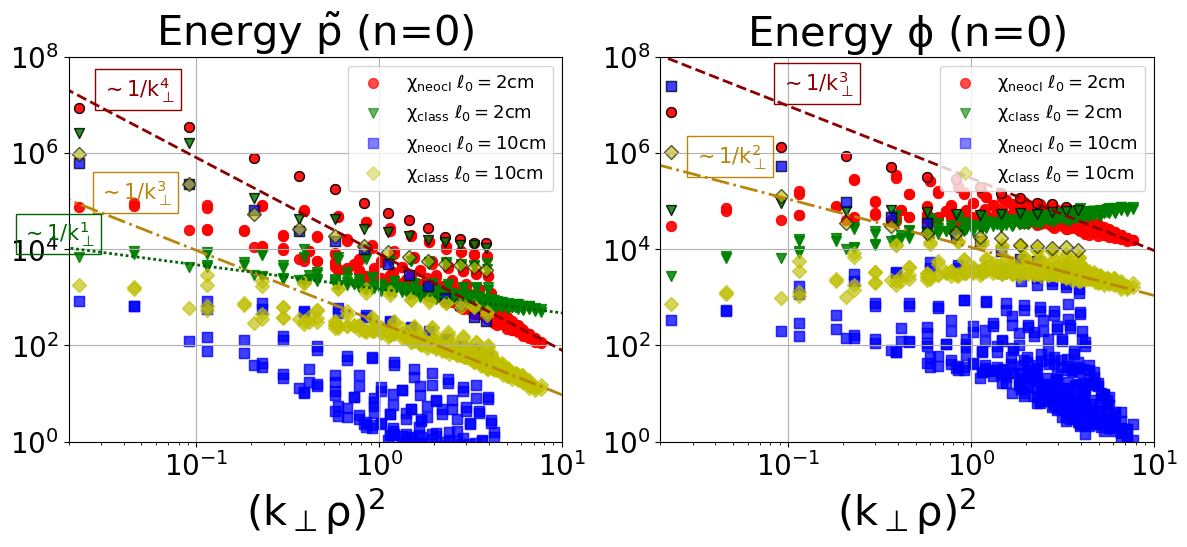}
\caption{The scatter plots of normalized axisymmetric spectral energies $E_{\tilde{p}}$ (left) and $E_\phi$ (right) vs $(k_\bot\rho)^2$ for the four considered cases with constant poloidal magnetic field and fully periodic BC. They are computed by averaging over a time window of $\sim 0.13ms$. Black edges are for those points corresponding to $m=0$ modes ($k_y=0$). For reference some curves are also plotted that approximate quite well some of the spectra profiles in SLAB (see Figs.\ref{en_spectra_SLAB} and \ref{en_spectra_NEW_SLAB}) and which behave as: $1/k_\bot^4$ (dashed dark red line), $1/k_\bot^3$ (dash-dotted dark yellow line) and $1/k_\bot$ (dotted dark green line) for $E_{\tilde{p}}$ (left) and $1/k_\bot^3$ (dashed dark red line) and $1/k_\bot^2$ (dash-dotted dark yellow line) for $E_\phi$ (right).}
\label{en_spectra_AC}
\end{figure}

\begin{itemize}
\item the plots of the fraction of $n=0$ (solid lines) and $n=m=0$ (dash-dotted lines) mode amplitudes in Fig.\ref{modes_AC}, that overlap in all cases except b) (green), where the latter is slightly lower than the former, signaling that most of the amplitude of toroidally symmetric mode is also poloidally symmetric,    
\item the 2D spectra in Fig.\ref{en_spectra_AC}, that show the anisotropy in $(k_x,k_y)$ plane with leading poloidally symmetric $m=0$ modes (outlined with black edges). 
\end{itemize}

The spectral energy values are now more scattered at given $k_\bot$ with respect to SLAB (see also the reference red, green and yellow curves in Fig.\ref{en_spectra_AC} that helps the comparison with the spectra in Figs.\ref{en_spectra_SLAB} and \ref{en_spectra_NEW_SLAB}), signaling strong anisotropy. Therefore the spectra are now quite different from those in SLAB, outlining how the modification of the drift terms according to Eq.(\ref{NablaparAC}) has a strong impact on the resulting spectral features. 

\begin{figure}[h]
\centering
\includegraphics[width=\textwidth]{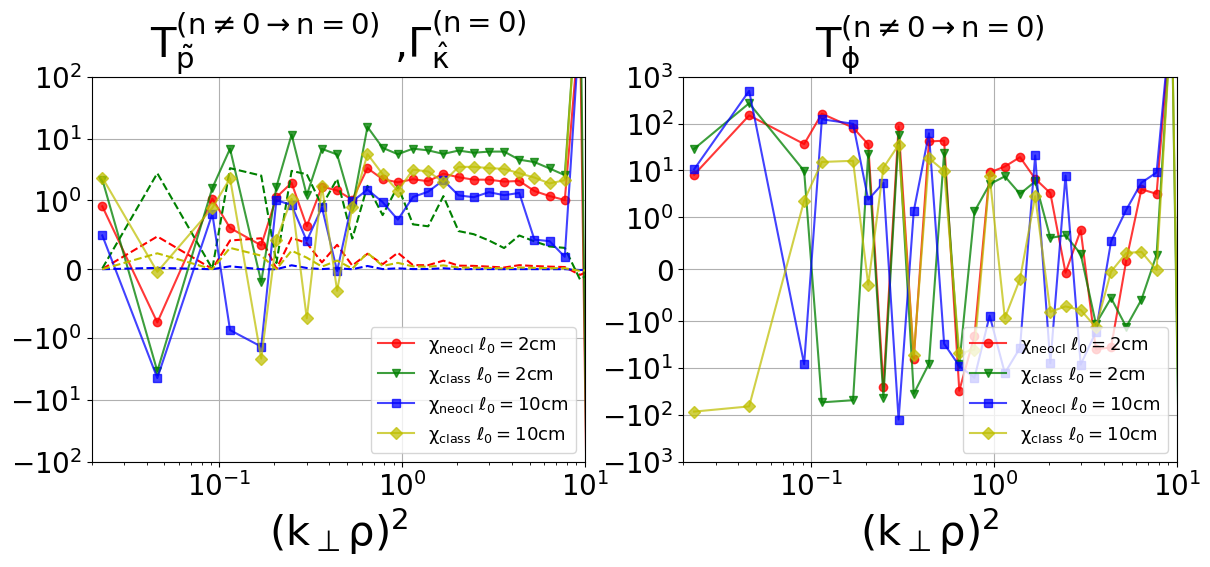}
\caption{The nonlinear transfer functions of toroidal transitions $n\neq0\rightarrow n=0$ (solid lines) for $\tilde{p}$ (left) and $\phi$ (right) vs $(k_\bot\rho)^2$ and the flux due to background pressure gradient (dashed lines, left) for the four considered cases with constant poloidal magnetic field and fully periodic BC. Each plotted value is normalized with the corresponding dissipative flux, {\it i.e.} the diffusivity and the viscous contributions for $\tilde{p}$ and $\phi$, respectively. The chosen values of $(k_\bot\rho)^2$ are constructed from the computational Cartesian grid in FS by taking the five smallest values of $k_\bot$ and by logarithmically sampling the relic part.}
\label{Tpphi_AC}
\end{figure}

Toroidal transition functions are plotted for logarithmically sampled $k_\bot$ values in Figs.\ref{Tpphi_AC}.

\begin{figure}[h]
\centering
\includegraphics[width=\textwidth]{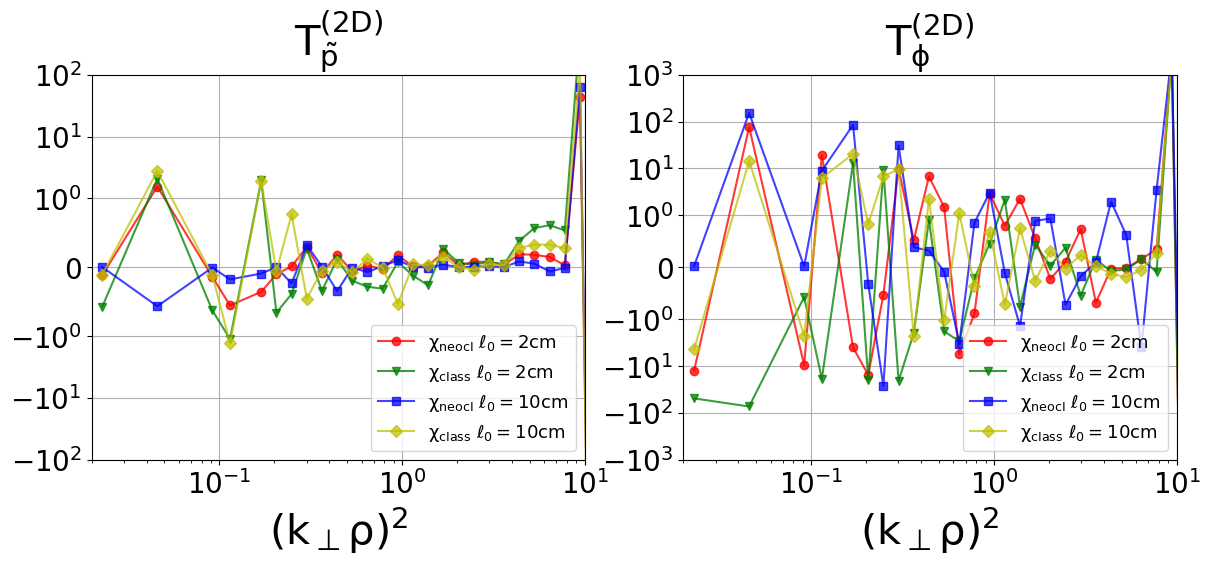}
\caption{The nonlinear transfer functions of 2D transitions $n=0\rightarrow n=0$ (solid lines) for $\tilde{p}$ (left) and $\phi$ (right) vs $(k_\bot\rho)^2$ for the four considered cases with constant poloidal magnetic field and fully periodic BC. Each plotted value is normalized with the corresponding dissipative flux, {\it i.e.} the diffusivity and the viscous contributions for $\tilde{p}$ and $\phi$, respectively.}
\label{T2D_AC}
\end{figure}

\begin{figure}[h]
\centering
\includegraphics[width=\textwidth]{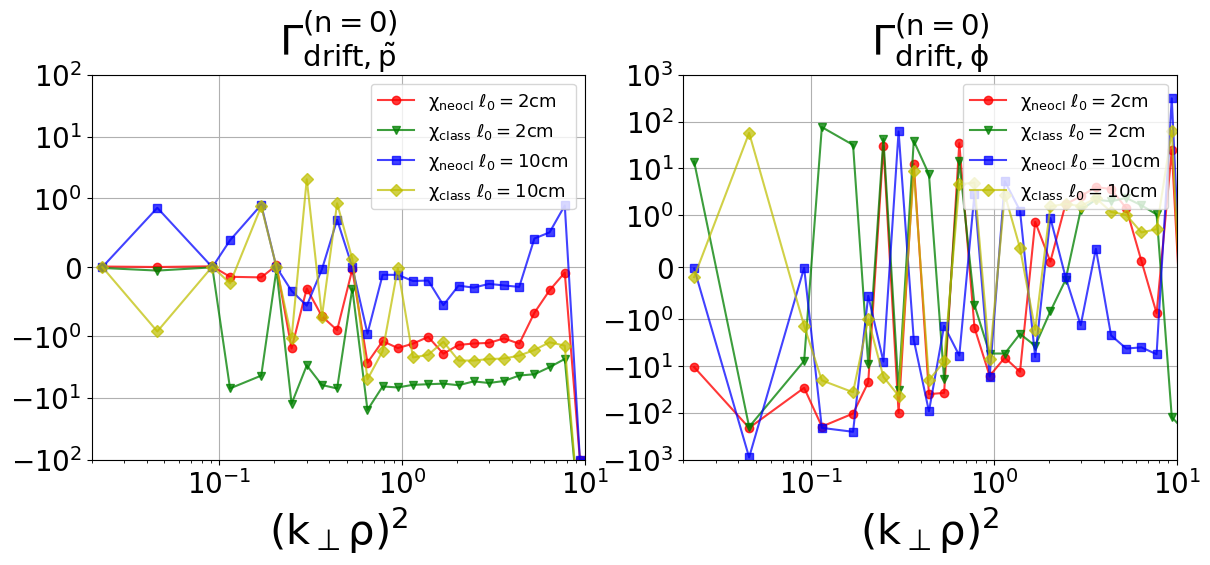}
\caption{The fluxes due to the drift coupling (solid lines) for $\tilde{p}$ (left) and $\phi$ (right) vs $(k_\bot\rho)^2$ for the four considered cases with constant poloidal magnetic field and fully periodic BC. Each plotted value is normalized with the corresponding dissipative flux, {\it i.e.} the diffusivity and the viscous contributions for $\tilde{p}$ and $\phi$, respectively.}
\label{Drift_AC}
\end{figure}

It is worth noting how $T^{(n\neq0\rightarrow n=0)}_{\tilde{p}}$ (left) are mostly positive, signaling an inverse toroidal cascade for $\tilde{p}$. Moreover, except for the largest wave-lengths ($(k_\bot \rho)^2<10^{-1}$), it is the leading source contribution with larger magnitude with respect to both $\Gamma_{\hat{k}}$ (dashed curves) and $T^{(n=0)}_{..}(2D)$ in Fig.\ref{T2D_AC} and it compensated by the drift term in Fig.\ref{Drift_AC} (left) .  

The interplay between toroidal and 2D transitions and the drift term is less clear for $\phi$, with $T^{(2D)}_{\phi}$ sub-dominant with respect to the to other contributions for short wave-length $(k_\bot\rho)>1$.

\begin{figure}[h]
\centering
\includegraphics[width=.7\textwidth]{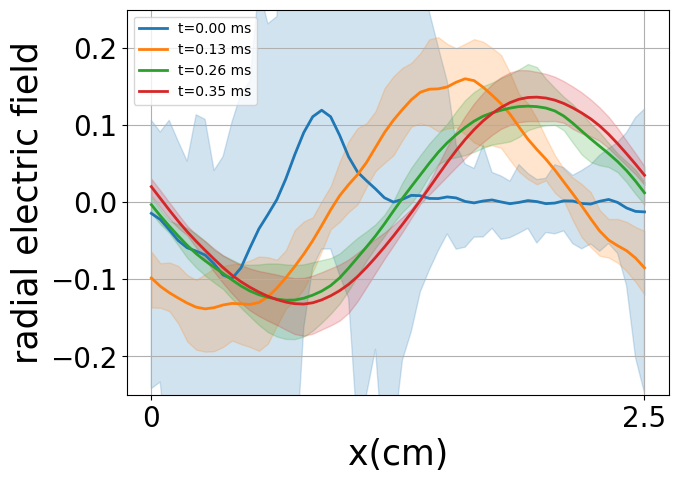}
\caption{The profile of the radial electric field in case c) with constant poloidal magnetic field and fully periodic BC at four given time values: start (blue), $t=0.04ms$ (orange), $t=0.09ms$ (green) and $t=0.17ms$ (red). The solid lines denote the averaged value along $y$ and $z$ directions, while the corresponding shaded regions cover the whole range of variability along such directions at the given time values.}
\label{rad_elf_AC}
\end{figure}

Case c) (blue) still exhibits the same kind of turbulence reduction in time shown in SLAB. In particular, a stationary sheared radial electric field develops (see Fig.\ref{rad_elf_AC}) with averaged value close to $l=1$ mode and small variations along poloidal and toroidal directions (small shaded region).

\begin{table}
\begin{tabular}{|c|c|}

      \hline \\
            \includegraphics[trim={3.5cm 1.5cm 3.5cm 1.7cm},clip,width=80mm]{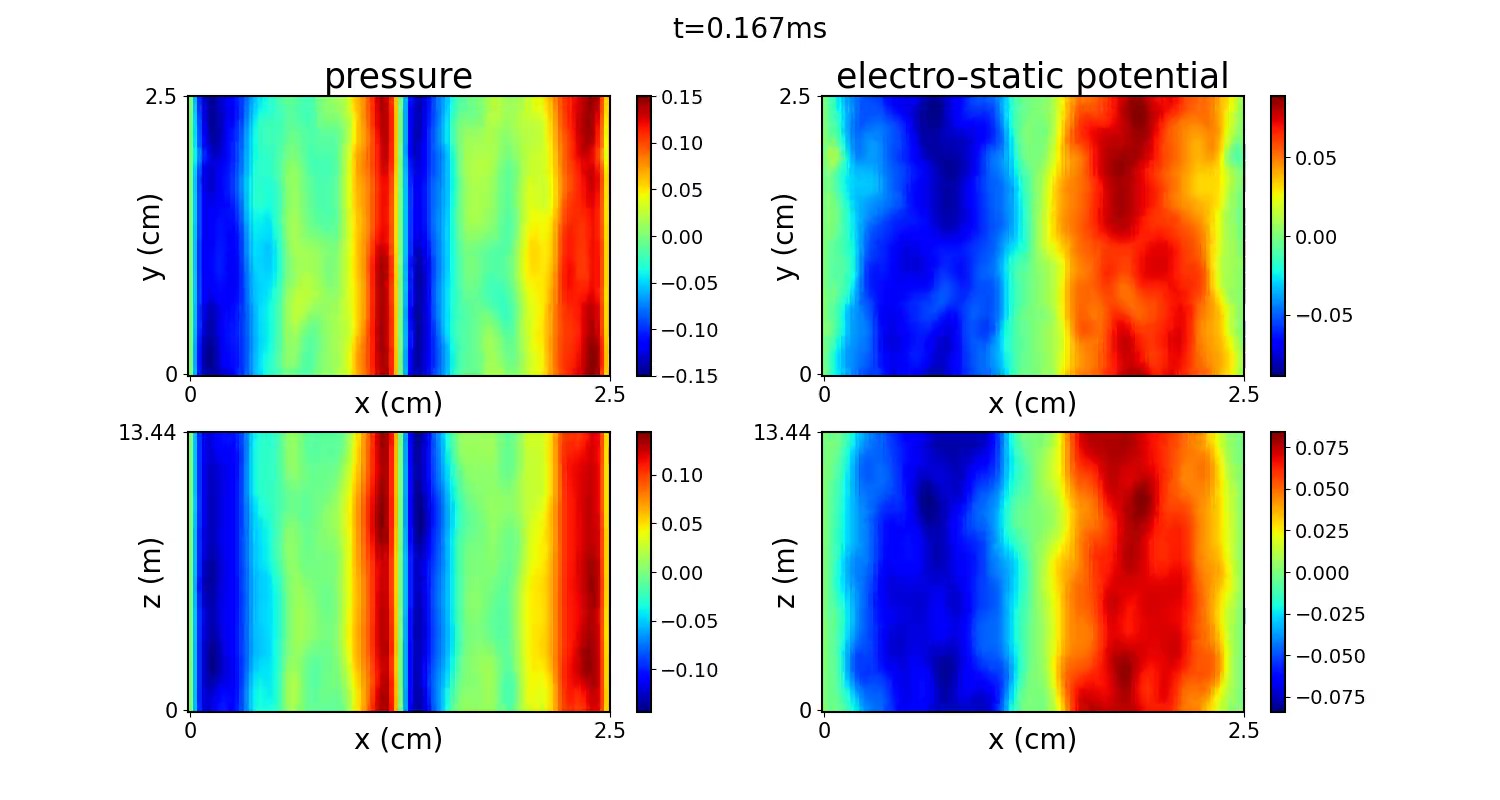} & \includegraphics[trim={3.5cm 1.5cm 3.5cm 1.7cm},clip,width=80mm]{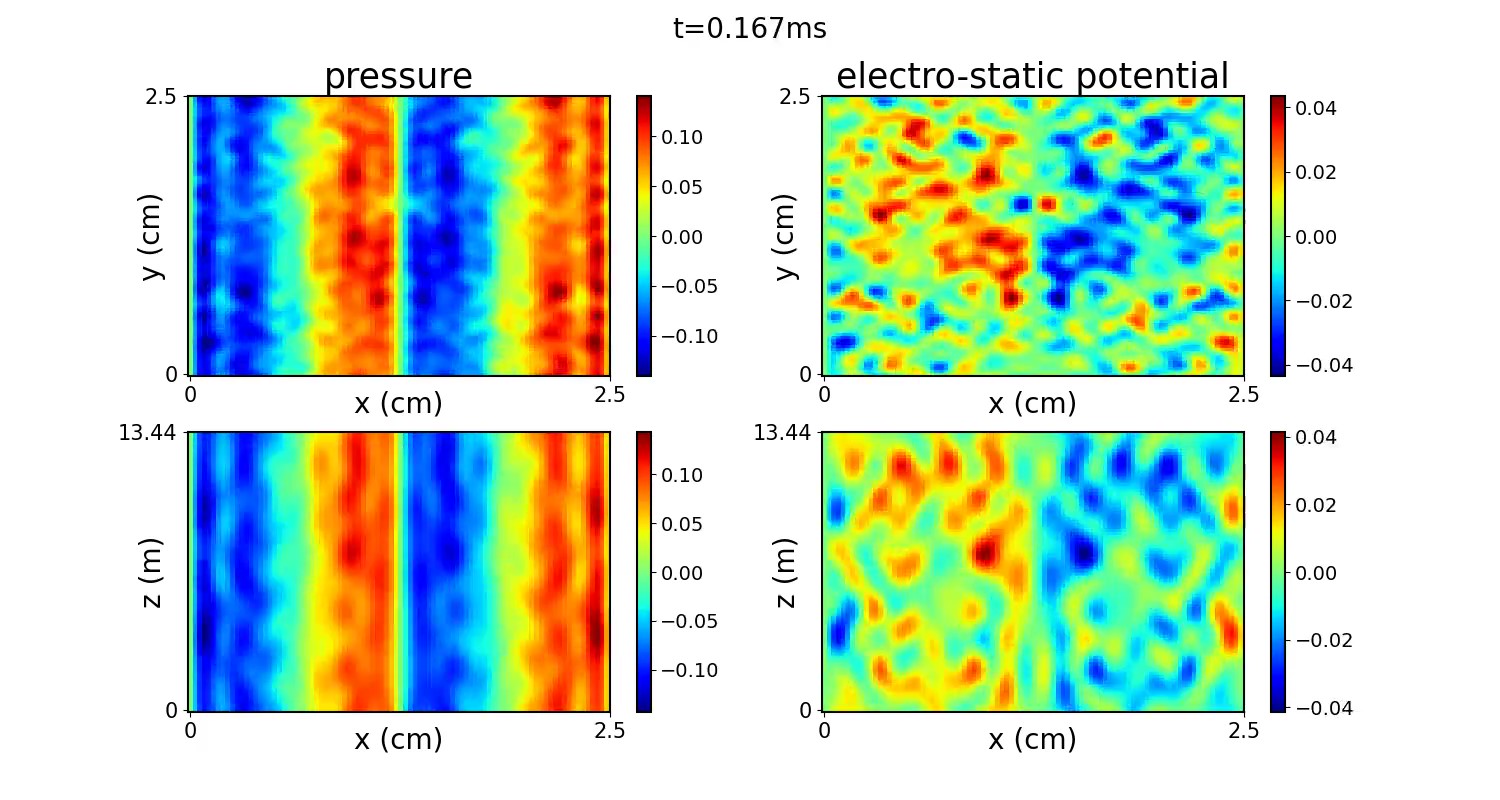} \\
			{\color{red}\textbullet} a) $\chi_{\mathrm{neocl}}$ $\ell_0$=2cm  & {\color{ForestGreen}\textbullet} b) $\chi_{\mathrm{class}}$ $\ell_0$=2cm \\
			\hline \\
      \includegraphics[trim={3.5cm 1.5cm 3.5cm 1.7cm},clip,width=80mm]{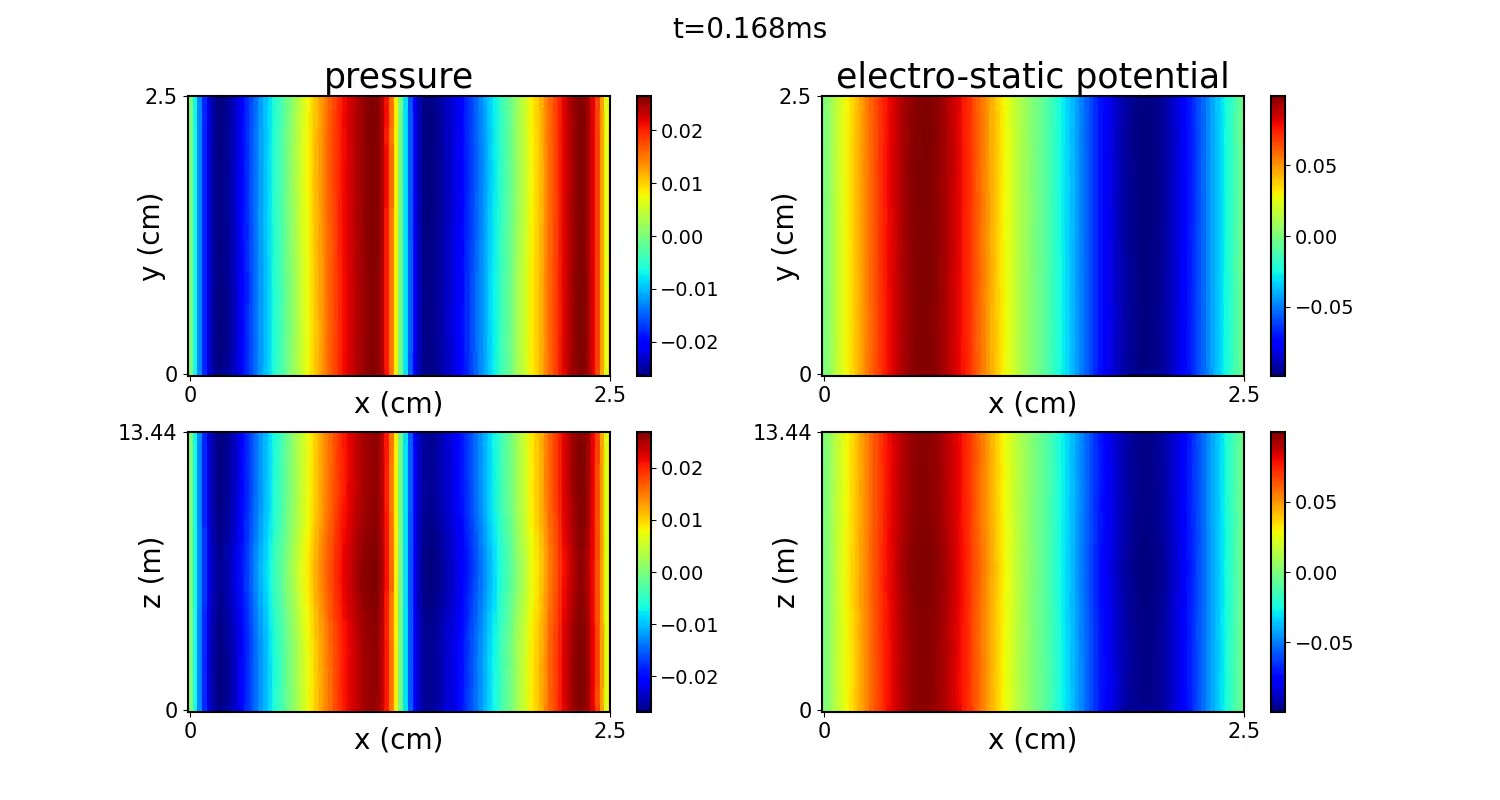} & \includegraphics[trim={3.5cm 1.5cm 3.5cm 1.7cm},clip,width=80mm]{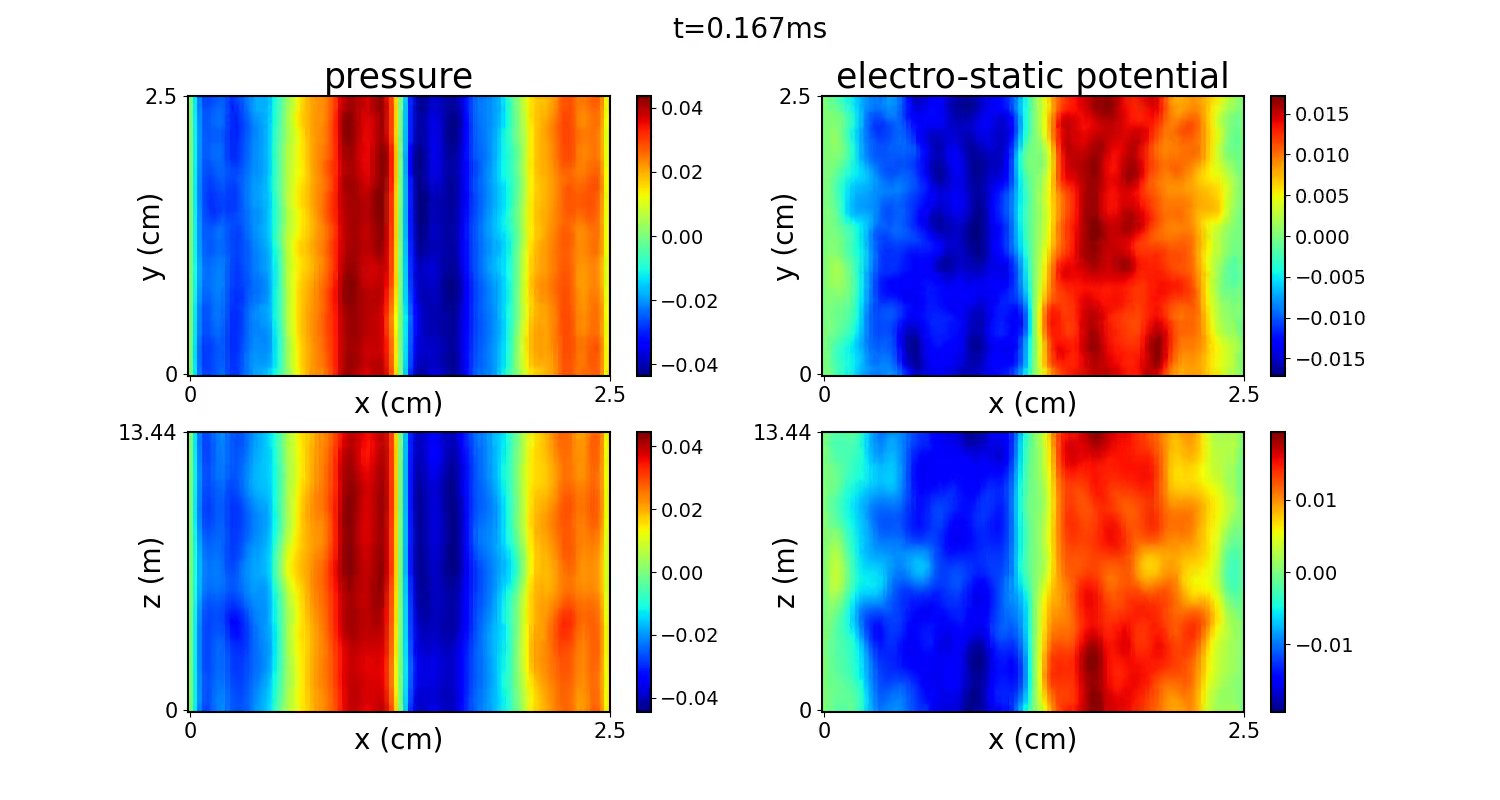} \\
			{\color{blue}\textbullet} c) $\chi_{\mathrm{neocl}}$ $\ell_0$=10cm & {\color{Goldenrod}\textbullet} d) $\chi_{\mathrm{class}}$ $\ell_0$=10cm \\			
			\hline 
\end{tabular}
\caption{The simulation results at time $t=0.167ms$ for constant poloidal magnetic field and vanishing radial Dirichlet BC. In each panel the figures above contains the poloidal sections, $(x,y)$ plane, of $\tilde{p}$ (left) and $\phi$ (right) while the figures below are for the radial-toroidal $(x,z)$ plane. The corresponding colorbars are shown next to each plot.}
\label{frames_NEW_AC}
\end{table}

\subsubsection{Dirichlet BC}

The simulation results with constant poloidal field and vanishing radial Dirichlet BC are shown in multi-panel figure Tab.\ref{frames_NEW_AC}. 

\begin{figure}[h]
\centering
\includegraphics[width=\textwidth]{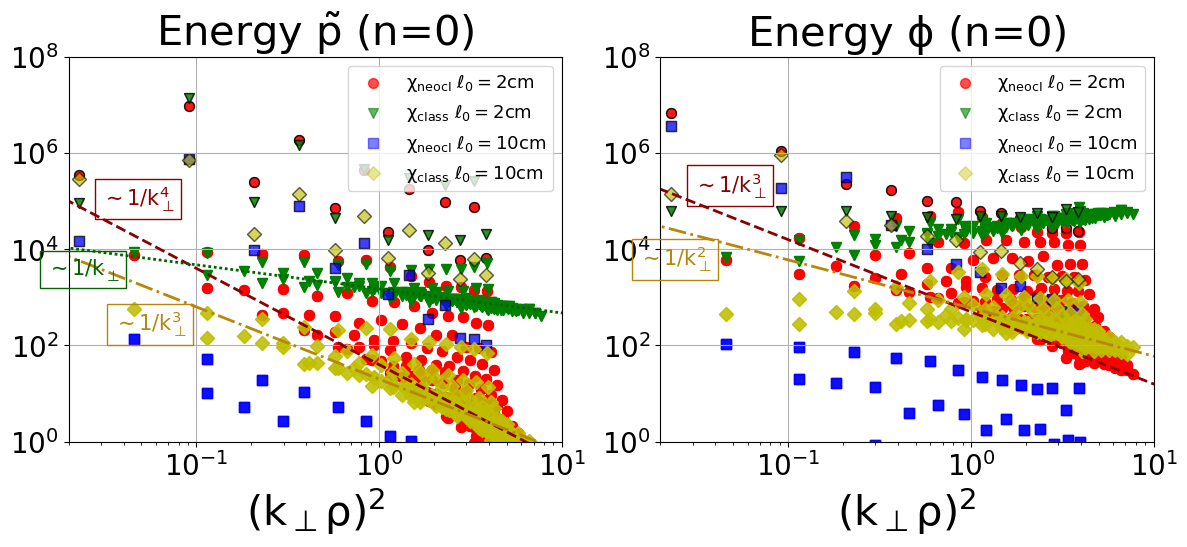}
\caption{The scatter plots of normalized axisymmetric spectral energies $E_{\tilde{p}}$ (left) and $E_\phi$ (right) vs $(k_\bot\rho)^2$ for the four considered cases with constant poloidal magnetic field and vanishing radial Dirichlet BC. They are computed by averaging over a time window of $\sim 0.13ms$. Black edges are for those points corresponding to $m=0$ modes ($k_y=0$). For reference some curves are also plotted that approximate quite well some of the spectra profiles in SLAB (see Figs.\ref{en_spectra_SLAB} and \ref{en_spectra_NEW_SLAB}) and which behave as: $1/k_\bot^4$ (dashed dark red line), $1/k_\bot^3$ (dash-dotted dark yellow line) and $1/k_\bot$ (dotted dark green line) for $E_{\tilde{p}}$ (left) and $1/k_\bot^3$ (dashed dark red line) and $1/k_\bot^2$ (dash-dotted dark yellow line) for $E_\phi$ (right).}
\label{en_spectra_NEW_AC}
\end{figure}

The specularity of $\tilde{p}$ and $\phi$ profiles along $x$ signals that the BC have been implemented correctly. The formation of more or less symmetric profiles along $y$ and $z$ depending on the considered case (c) the most symmetric, b) the least symmetric ones) is analogous to the results with fully periodic BC. The spectral anisotropy, that in SLAB resulted from the implementation of vanishing radial Dirichlet BC, sums up to that due to the poloidal drift component. In fact, the resulting energy spectra in Fig.\ref{en_spectra_NEW_AC} present even more anisotropy with respect to those in Fig.\ref{en_spectra_AC}, especially one can notice a major enhancement of $m=0$ modes for $\tilde{p}$.

The analysis of 3D and 2D transitions and of the drift terms provide similar qualitative results as in the case of fully periodic BC, with inverse toroidal cascade for $\tilde{p}$ and negligible 2D transitions at long wave-lengths for $\phi$. Case c) (blue) still shows turbulence reduction in time and the formation of a stationary radial electric field. 

Therefore, the simulations with constant poloidal magnetic field outline how the introduction of a poloidal magnetic field component provides an inverse poloidal cascade in addition to the toroidal one already seen in SLAB. This fact suggests that the direction of the inverse cascade in FS is determined by that of the drift term. The resulting 2D spectra are anisotropic with leading poloidally symmetric modes and their profiles are affected by the value of the dissipative ($\chi_\bot$) and source ($\ell_0$) parameters similarly to SLAB case, {\it i.e} they get wider by reducing $\chi_\bot$ and/or $\ell_0$.

\subsection{Radial magnetic shear}\label{sec3.3}
A radially sheared poloidal field is discussed in this section, with magnetic components $B_x=0$ and $B_y$ ranging from SLAB case at the 
midpoint $l_x/2$ of the radial simulation domain $[0,l_x]$ to $\pm$ the constant poloidal field discussed in the previous section at the borders $x=0$ and $x=l_x$. The chosen form is 
\begin{equation}
B_y=B_z\frac{2\alpha}{l_x}\left(x-\frac{l_x}{2}\right)\,\label{shmg}
\end{equation}
with the parameter $\alpha=\frac{L_\bot}{L_\parallel}=1.5\cdot 10^{-3}$. The expressions of $\Delta_\bot$ and $\nabla_\parallel$ are formally the same as in the case with constant poloidal field, Eqs.(\ref{DeltaperpAC}) and (\ref{NablaparAC}), respectively, but since $\hat{b}_y$ is now a function of $x$ radial coordinate the corresponding expressions are not anymore multiplicative operators in FS and are numerically computed with the pseudospectral method.  

The magnetic shear is known to stabilize turbulence and this is confirmed by the simulations shown in the multi-panel figure Tab.\ref{frames_shear}, which generically exhibit weaker turbulence amplitude with respect to the analogous results in the two other considered magnetic configurations (shown in the multi-panel figures Tabs.\ref{frames_SLAB} and \ref{frames_AC}). 

\begin{table}
\begin{tabular}{|c|c|}

      \hline \\
            \includegraphics[trim={3.5cm 1.5cm 3.5cm 1.7cm},clip,width=80mm]{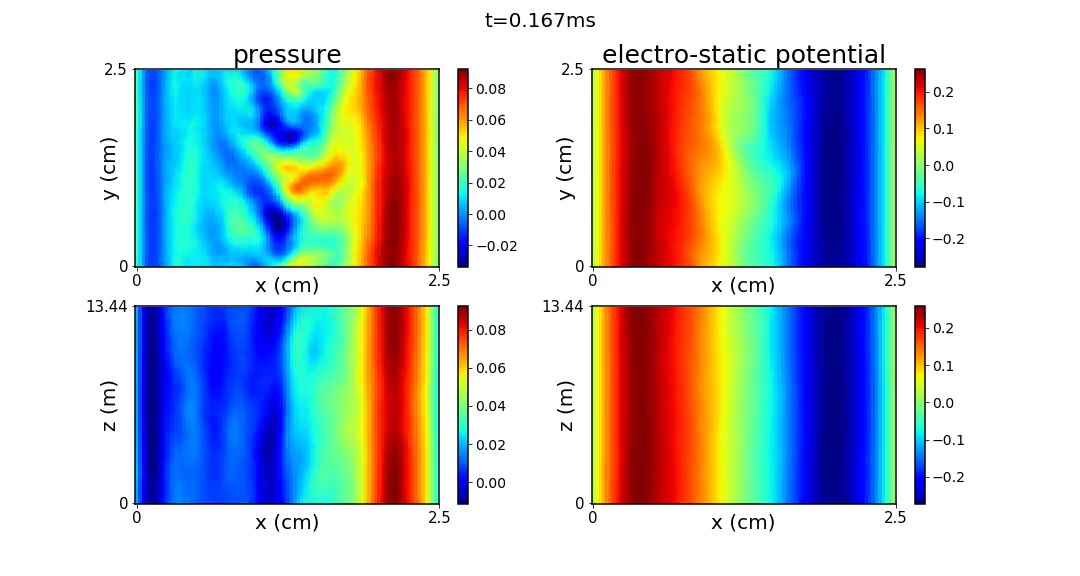} & \includegraphics[trim={3.5cm 1.5cm 3.5cm 1.7cm},clip,width=80mm]{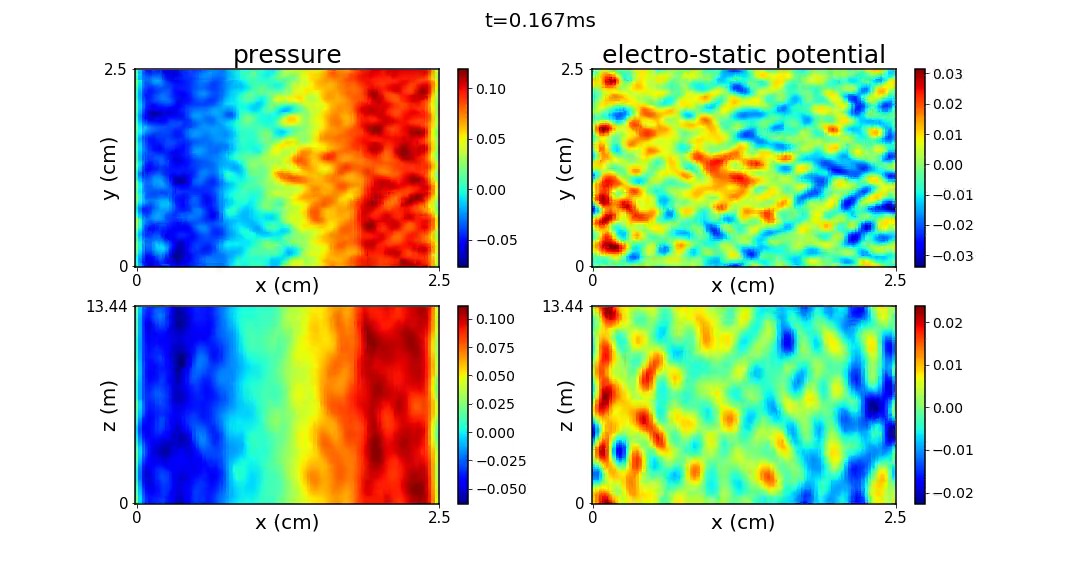} \\
			{\color{red}\textbullet} a) $\chi_{\mathrm{neocl}}$ $\ell_0$=2cm  & {\color{ForestGreen}\textbullet} b) $\chi_{\mathrm{class}}$ $\ell_0$=2cm \\
			\hline \\
      \includegraphics[trim={3.5cm 1.5cm 3.5cm 1.7cm},clip,width=80mm]{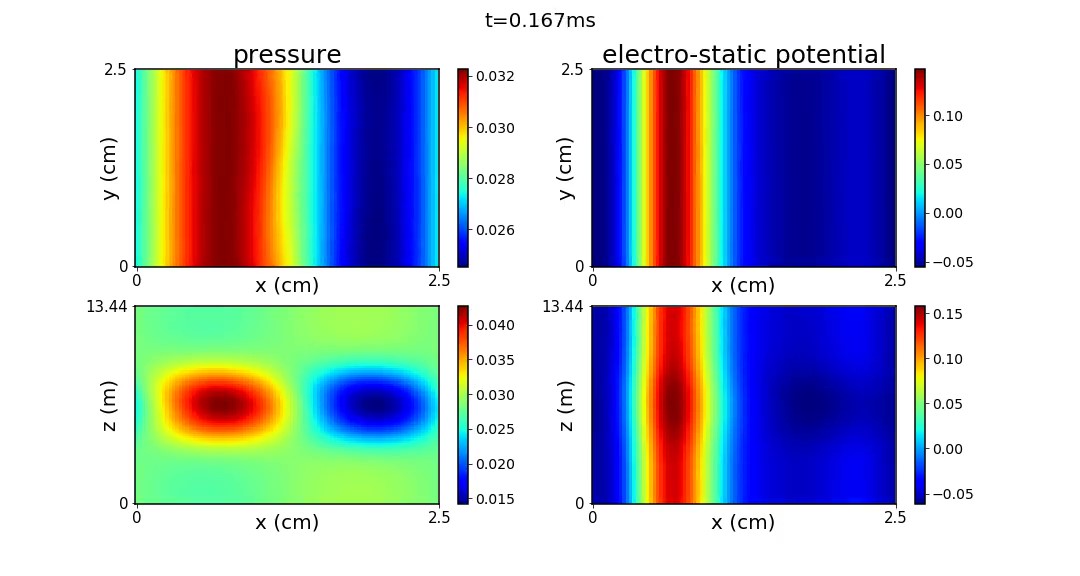} & \includegraphics[trim={3.5cm 1.5cm 3.5cm 1.7cm},clip,width=80mm]{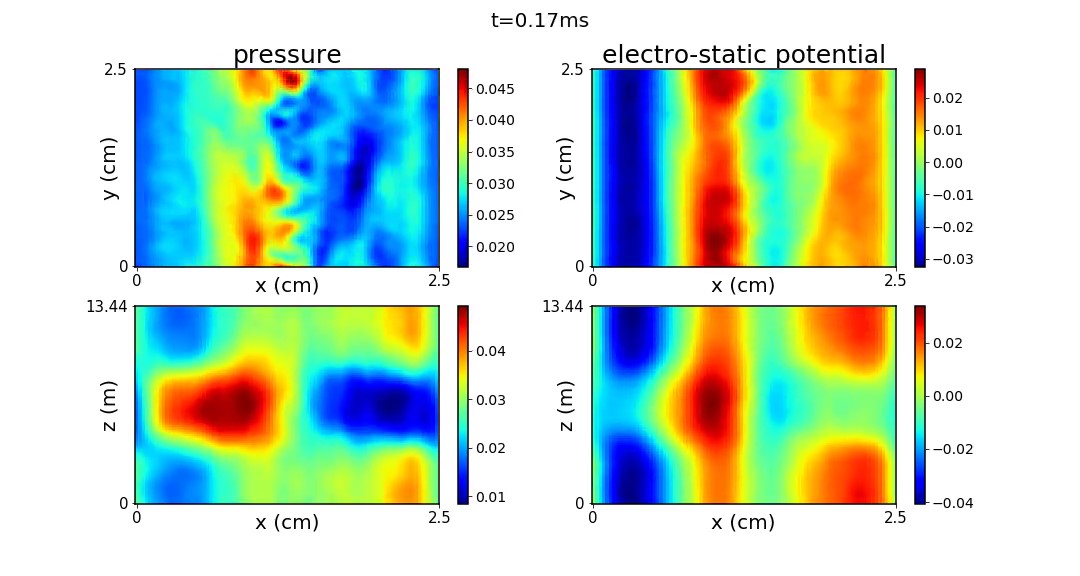} \\
			{\color{blue}\textbullet} c) $\chi_{\mathrm{neocl}}$ $\ell_0$=10cm & {\color{Goldenrod}\textbullet} d) $\chi_{\mathrm{class}}$ $\ell_0$=10cm \\			
			\hline 
\end{tabular}
\caption{The simulation results at time $t=0.167ms$ for radially sheared poloidal magnetic field and fully periodic BC. In each panel the figures above contain the poloidal sections, $(x,y)$ plane, of normalized $\tilde{p}$ (left) and $\phi$ (right) while the figures below for the radial-toroidal $(x,z)$ plane. The corresponding colorbars are shown next to each plot.}
\label{frames_shear}
\end{table}

\begin{figure}[h]
\centering
\includegraphics[width=\textwidth]{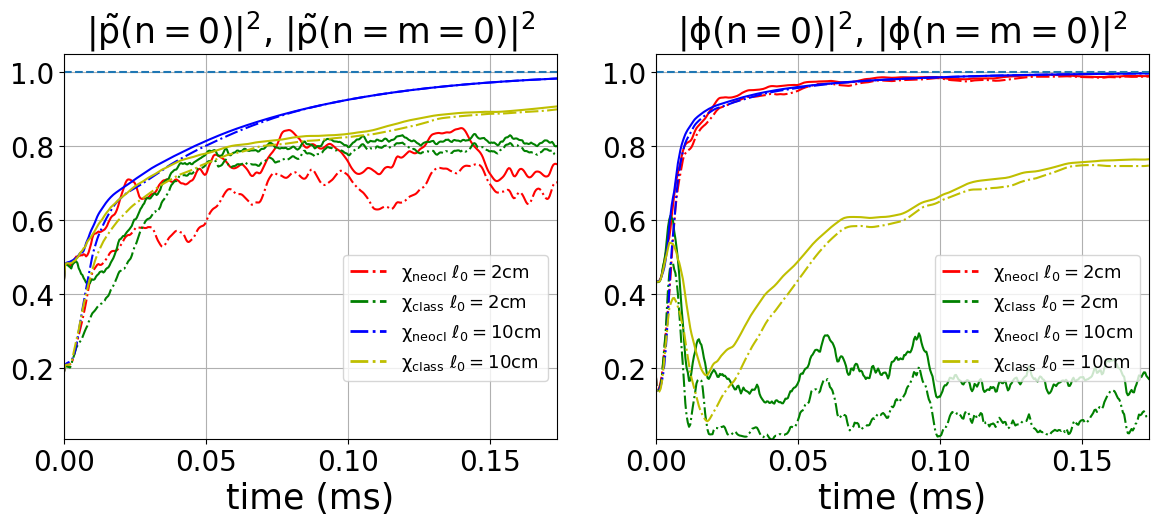}
\caption{The ratio of $n=0$ (solid lines) and of $n=m=0$ (dash-dotted lines) mode amplitudes over the total amplitude for $\tilde{p}$ (left) and $\phi$ (right) vs time (in ms) for the four considered cases with radially sheared poloidal magnetic field and fully periodic BC.}
\label{modes_shear}
\end{figure}

The fraction of $n=0$ (solid lines) and $n=m=0$ (dash-dotted lines) mode amplitudes are plotted in Fig.\ref{modes_shear}. The solid lines for $\tilde{p}$ approach $\sim 0.8$ for cases a) and b) (red and green), $\sim 0.9$ for case d) (yellow) and $\sim 1$ for c) (blue), while the same curves for $\phi$ reach $1$ in the two neoclassical cases a) and c) (red and blue), $\sim 0.75$ for d) (yellow) and $0.2$ for b) (green). The dash-dotted lines are slightly lower than solid ones, meaning that most of the toroidally symmetric modes are poloidally symmetric too. Therefore, we get a similar cascade along $y$ as for a constant poloidal field (see Fig.\ref{modes_AC}), while the toroidal cascade more resembles the SLAB profile (see Fig.\ref{modes_SLAB}).      

\begin{figure}[h]
\centering
\includegraphics[width=\textwidth]{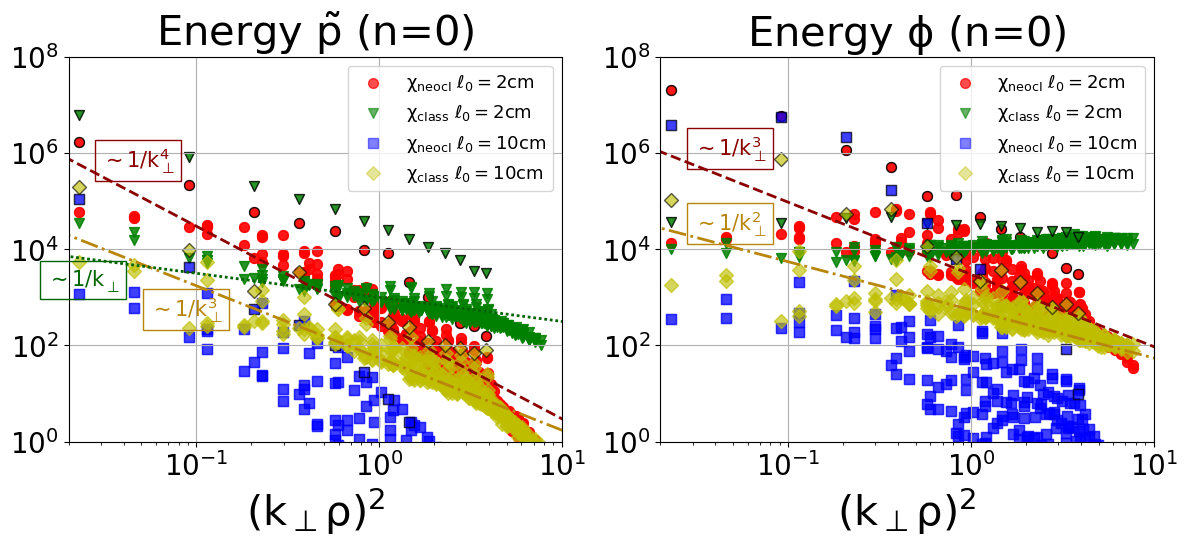}
\caption{The scatter plots of normalized axisymmetric spectral energies $E_{\tilde{p}}$ (left) and $E_\phi$ (right) vs $(k_\bot\rho)^2$ for the four considered cases with radially sheared poloidal magnetic field and fully periodic BC. They are computed by averaging over a time window of $\sim 0.13ms$. Black edges are for those points corresponding to $m=0$ modes ($k_y=0$). For reference some curves are also plotted that approximate quite well some of the spectra profiles in SLAB (see Figs.\ref{en_spectra_SLAB} and \ref{en_spectra_NEW_SLAB}) and which behave as: $1/k_\bot^4$ (dashed dark red line), $1/k_\bot^3$ (dash-dotted dark yellow line) and $1/k_\bot$ (dotted dark green line) for $E_{\tilde{p}}$ (left) and $1/k_\bot^3$ (dashed dark red line) and $1/k_\bot^2$ (dash-dotted dark yellow line) for $E_\phi$ (right).}
\label{en_spectra_shear}
\end{figure}

The energy spectra are shown in Fig.\ref{en_spectra_shear} and they exhibit intermediate properties between the magnetic configurations previously discussed: a similar degree of anisotropy as for constant poloidal magnetic field in Fig.\ref{en_spectra_AC} and an averaged profile that roughly follows the reference curves reproducing the behavior in SLAB . 

\begin{figure}[h]
\centering
\includegraphics[width=\textwidth]{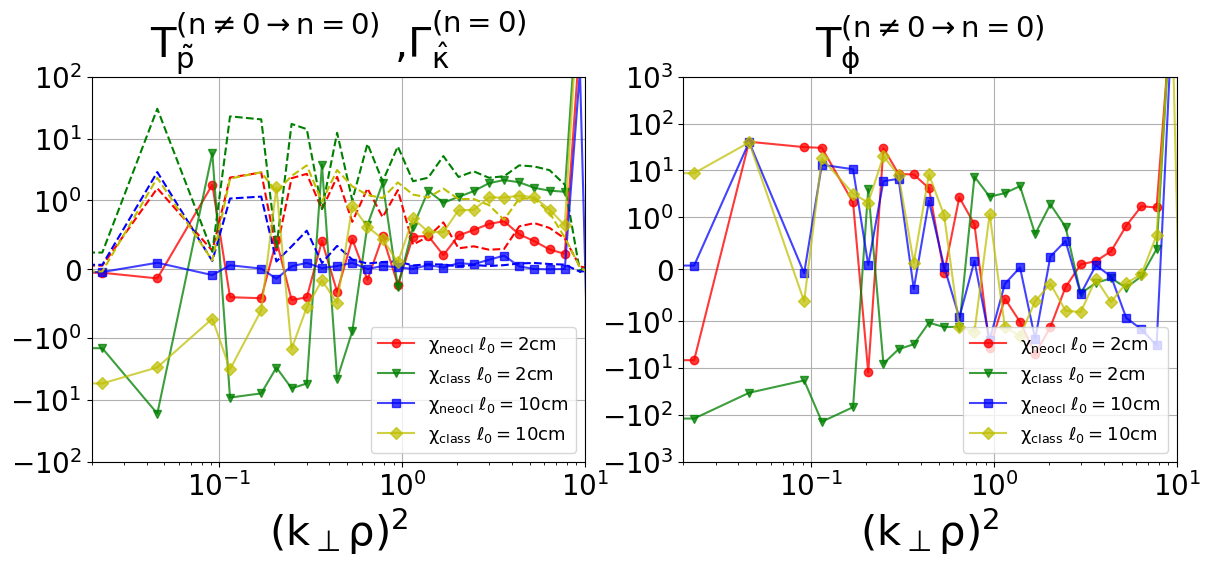}
\caption{The nonlinear transfer functions of toroidal transitions $n\neq0\rightarrow n=0$ (solid lines) for $\tilde{p}$ (left) and $\phi$ (right) vs $(k_\bot\rho)^2$ and the flux due to background pressure gradient (dashed lines, left) for the four considered cases with radially sheared poloidal magnetic field and fully periodic BC. Each plotted value is normalized with the corresponding dissipative flux, {\it i.e.} the diffusivity and the viscous contributions for $\tilde{p}$ and $\phi$, respectively. The chosen values of $(k_\bot\rho)^2$ are constructed from the computational Cartesian grid in FS by taking the five smallest values of $k_\bot$ and by logarithmically sampling the relic part.}
\label{Tpphi_shear}
\end{figure}

Toroidal transfer functions for $\tilde{p}$ in Fig.\ref{Tpphi_shear} (left) have a sort of mixed behavior: there is a direct toroidal cascade at short wave-lengths as in SLAB and an inverse one at long wave-lengths as for the constant poloidal field. These quantities are very close to zero in the two neoclassical cases a) and c) (red and blue) which correspond to toroidal and poloidal symmmetry (this can be seen in both Figs.\ref{frames_shear} and \ref{modes_shear}). 

\begin{figure}[h]
\centering
\includegraphics[width=\textwidth]{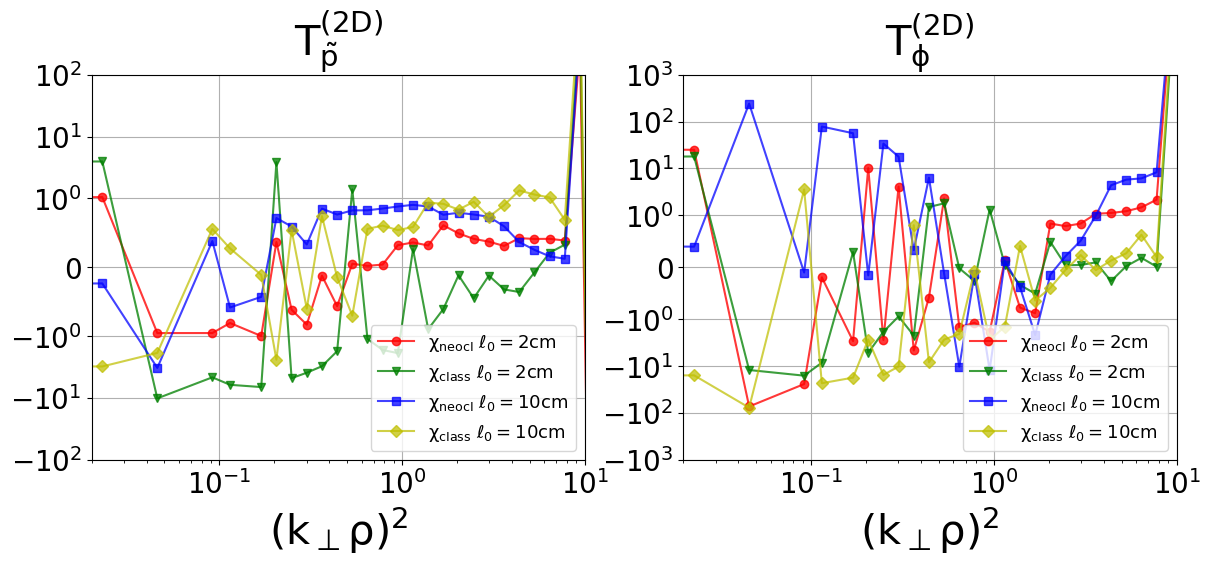}
\caption{The nonlinear transfer functions of 2D transitions $n=0\rightarrow n=0$ (solid lines) for $\tilde{p}$ (left) and $\phi$ (right) vs $(k_\bot\rho)^2$ for the four considered cases with radially sheared poloidal magnetic field and fully periodic BC. Each plotted value is normalized with the corresponding dissipative flux, {\it i.e.} the diffusivity and the viscous contributions for $\tilde{p}$ and $\phi$, respectively.}
\label{Tpphi2D_shear}
\end{figure}

The toroidal transfer functions for $\phi$ (right) have strong fluctuations for neighbor sampled $(k_\bot\rho)^2$ values and they strongly depend from case to case. Their magnitude are comparable to those of 2D transitions in Fig.\ref{Tpphi2D_shear}. 

\begin{figure}[h]
\centering
\includegraphics[width=\textwidth]{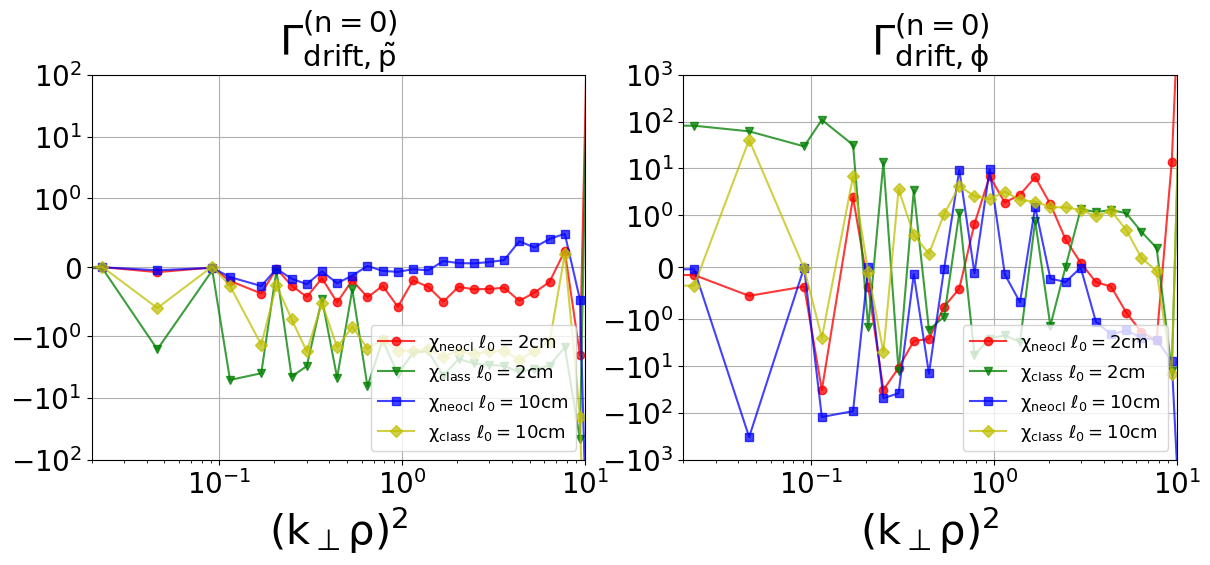}
\caption{The fluxes due to the drift coupling (solid lines) for $\tilde{p}$ (left) and $\phi$ (right) vs $(k_\bot\rho)^2$ for the four considered cases with radially sheared poloidal magnetic field and fully periodic BC. Each plotted value is normalized with the corresponding dissipative flux, {\it i.e.} the diffusivity and the viscous contributions for $\tilde{p}$ and $\phi$, respectively.}
\label{Drift_shear}
\end{figure}

Finally, the drift fluxes are shown in Fig.\ref{Drift_shear} and they outline similar features as for a constant poloidal field: they are mostly sink contributions for $\tilde{p}$ (negative), while no definite behavior is seen for $\phi$. In all cases, they have a similar magnitude as toroidal and 2D transitions.  

The conclusion of fluxes analysis is that spectral profiles are determined by the combination of both 2D and toroidal transitions with the linear contributions, {\it i.e.} those due to background pressure gradient for $\tilde{p}$, drift terms and dissipation. 

\begin{figure}[h]
\centering
\includegraphics[width=.7\textwidth]{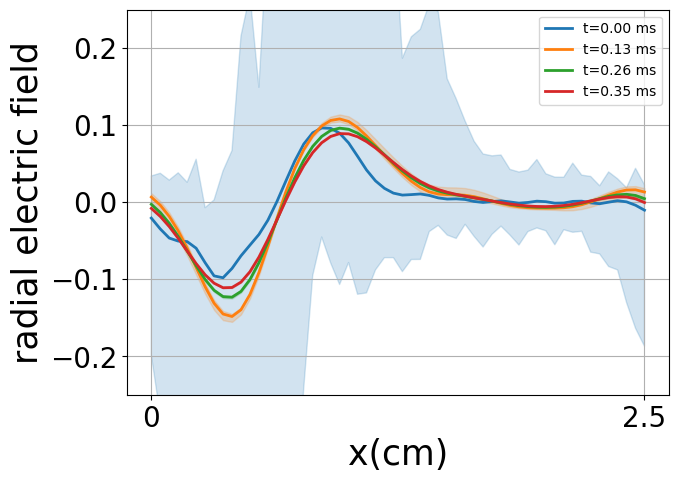}
\caption{The profile of the radial electric field in case c) with radially sheared poloidal magnetic field and fully periodic BC at four given time values: start (blue), $t=0.04ms$ (orange), $t=0.09ms$ (green) and $t=0.17ms$ (red). The solid lines denote the averaged value along $y$ and $z$ directions, while the corresponding shaded regions cover the whole range of variability along such directions at the given time values.}
\label{rad_elf_shear}
\end{figure}

Like in the other considered magnetic geometries, case c) is peculiar as turbulence gets suppressed in time and a sheared radial electric field is generated as shown in Fig.\ref{rad_elf_shear}. The profile is toroidally and poloidally symmetric, as the extent of the shaded area is very small since $t=0.13ms$ (orange) and it is nonvanishing only in the first half of $x$ simulation domain, where the inital wave-packet is localized.   

\subsubsection{Dirichlet BC}

The simulation results with the sheared magnetic geometry in Eq.(\ref{shmg}) and vanishing radial Dirichlet BC are shown in the multi-panel figure Tab.\ref{frames_NEW_shear}. 

\begin{table}
\begin{tabular}{|c|c|}
      \hline \\
            \includegraphics[trim={3.5cm 1.5cm 3.5cm 1.7cm},clip,width=80mm]{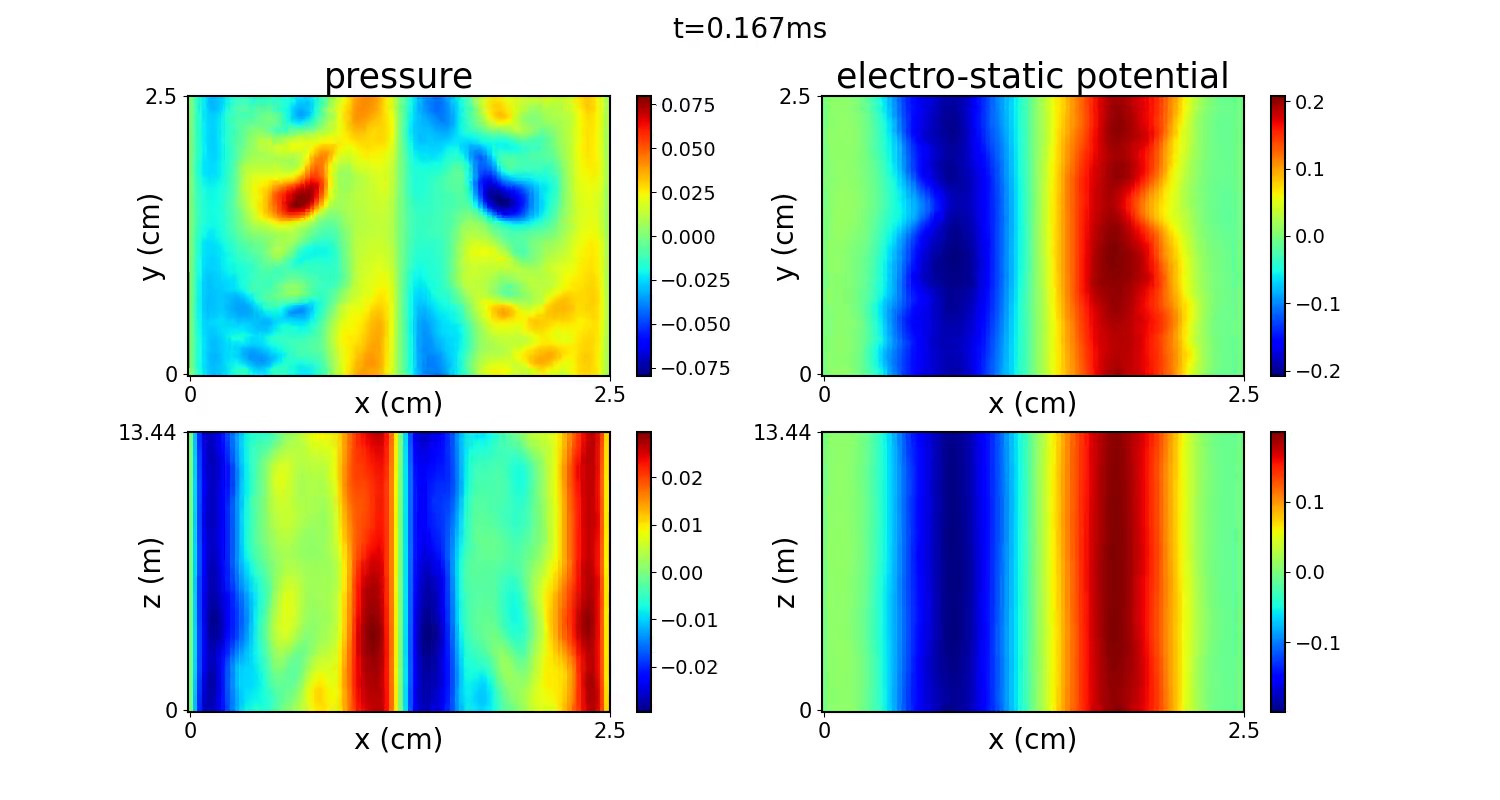} & \includegraphics[trim={3.5cm 1.5cm 3.5cm 1.7cm},clip,width=80mm]{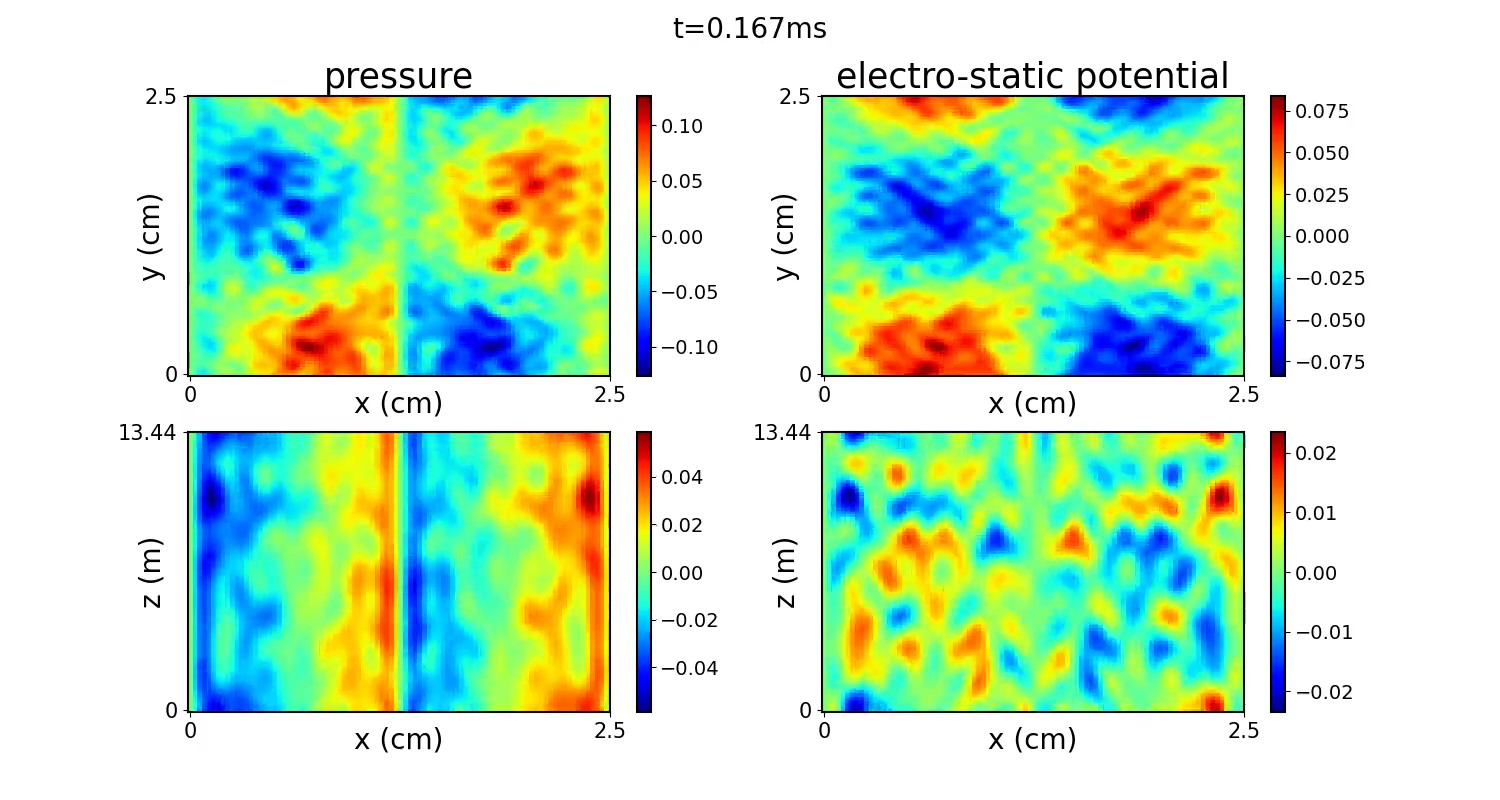} \\
			{\color{red}\textbullet} a) $\chi_{\mathrm{neocl}}$ $\ell_0$=2cm  & {\color{ForestGreen}\textbullet} b) $\chi_{\mathrm{class}}$ $\ell_0$=2cm \\
			\hline \\
      \includegraphics[trim={3.5cm 1.5cm 3.5cm 1.7cm},clip,width=80mm]{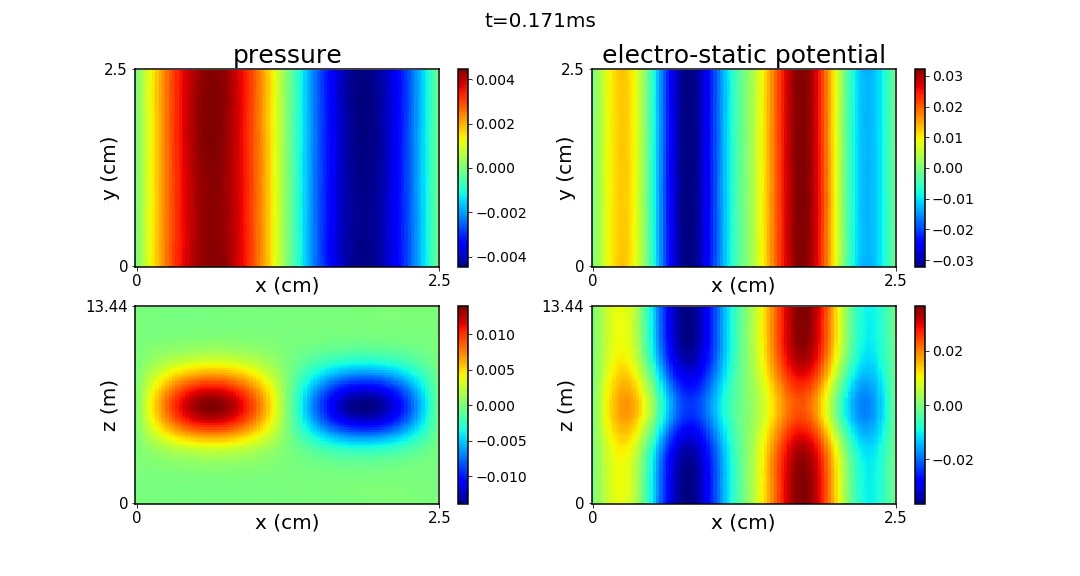} & \includegraphics[trim={3.5cm 1.5cm 3.5cm 1.7cm},clip,width=80mm]{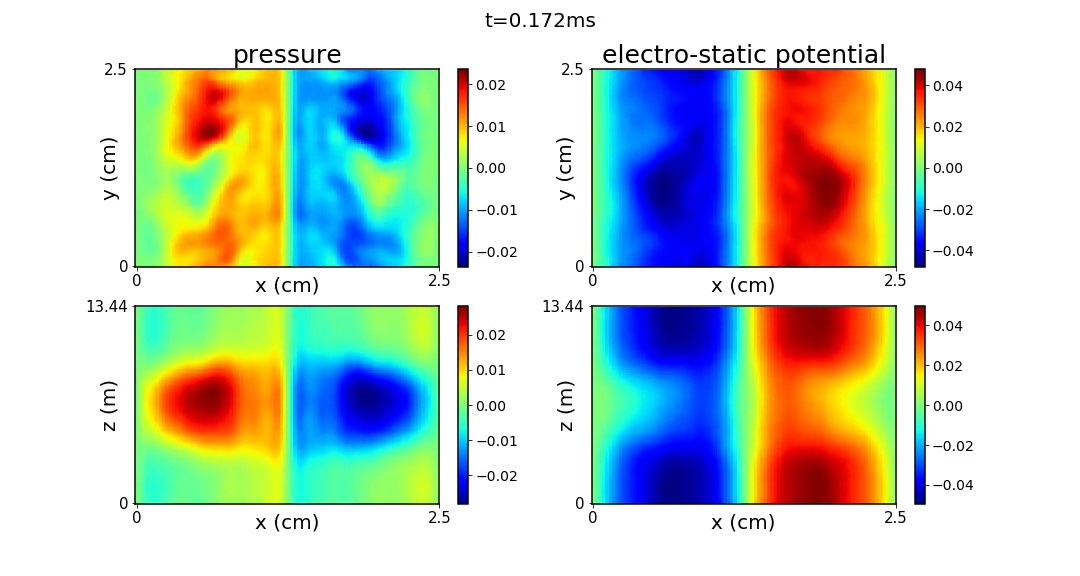} \\
			{\color{blue}\textbullet} c) $\chi_{\mathrm{neocl}}$ $\ell_0$=10cm & {\color{Goldenrod}\textbullet} d) $\chi_{\mathrm{class}}$ $\ell_0$=10cm \\			
			\hline 
\end{tabular}
\caption{The simulation results at time $t=0.167ms$ for radially sheared poloidal magnetic field and vanishing radial Dirichlet BC. In each panel the figures above contain the poloidal sections, $(x,y)$ plane, of normalized $\tilde{p}$ (left) and $\phi$ (right) while the figures below are for the radial-toroidal $(x,z)$ plane. The corresponding colorbars are shown next to each plot.}
\label{frames_NEW_shear}
\end{table}

\begin{figure}[h]
\centering
\includegraphics[width=\textwidth]{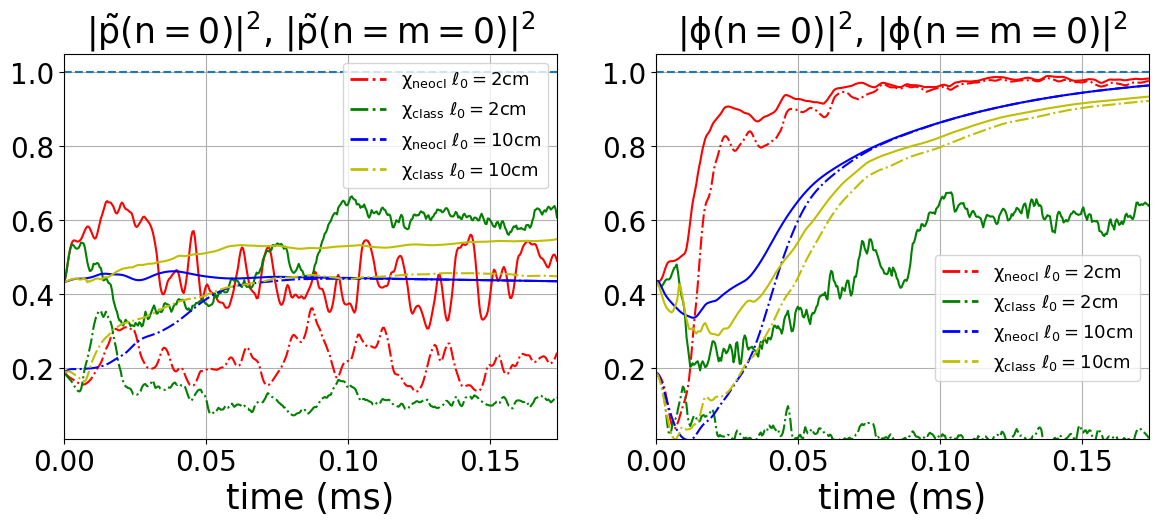}
\caption{The ratio of $n=0$ (solid lines) and of $n=m=0$ (dash-dotted lines) mode amplitudes over the total amplitude for $\tilde{p}$ (left) and $\phi$ (right) vs time (in ms) for the four considered cases with radially sheared poloidal magnetic field and vanishing radial Dirichlet BC.}
\label{modes_NEW_shear}
\end{figure}

It is worth noting how poloidally symmetric and, to a minor extent, toroidally symmetric configurations develop in $\phi$, especially in cases a), c) and d), while the pressure profiles keep a certain degree of asymmetry in both directions in all cases. This is confirmed by the plots in Fig.\ref{modes_NEW_shear} of the fraction of $n=0$ (solid lines) and $n=m=0$ modes (dash-dotted lines). For $\phi$ (right), in all cases except b) (blue) the solid and dot-dashed lines overlap and reach values close to one, signaling that both poloidally and toroidally symmetric configurations are obtained. This finding is in qualitative agreement with the case with fully periodic BC (see Fig.\ref{modes_NEW_shear}), while in case b) (green) one gets an enhancement of toroidally symmetric mode amplitude by a factor 3 (from $20\%$ in Fig.\ref{modes_NEW_shear} to $60\%$ here).

The same plots for $\tilde{p}$ (left) exhibit much lower degree of symmetry in both directions: the fraction of toroidally symmetric mode amplitude (solid lines) is in the range $40\%\div 60\%$ and except for c) (blue) the fraction of poloidally symmetric modes (dash-dotted lines) is significantly lower. 

\begin{figure}[h]
\centering
\includegraphics[width=\textwidth]{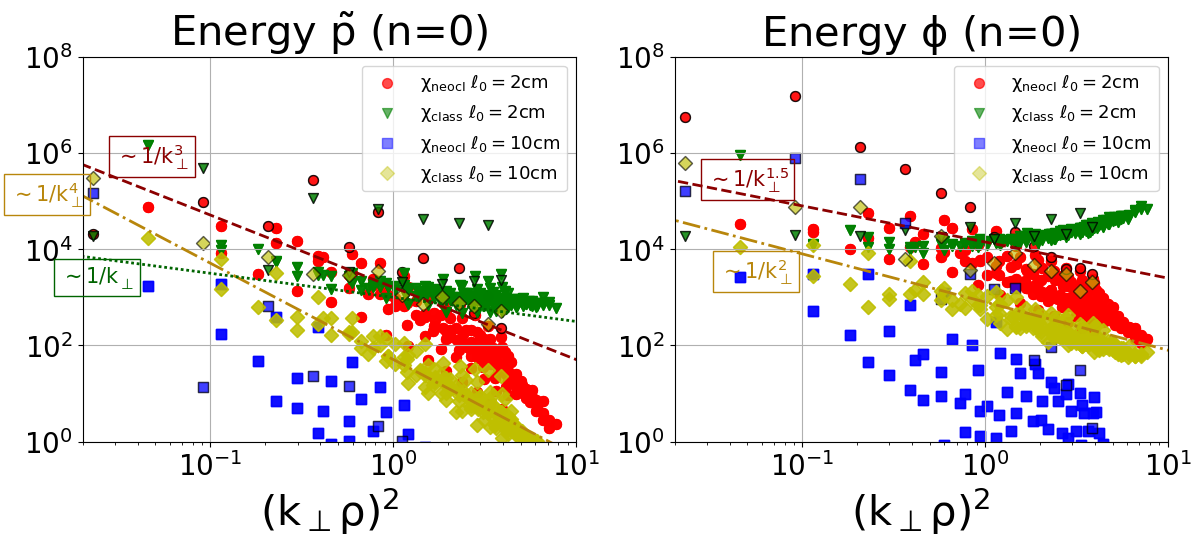}
\caption{The scatter plots of normalized axisymmetric spectral energies $E_{\tilde{p}}$ (left) and $E_\phi$ (right) vs $(k_\bot\rho)^2$ for the four considered cases with radially sheared poloidal magnetic field and vanishing radial Dirichlet BC. They are computed by averaging over a time window of $\sim 0.13ms$. Black edges are for those points corresponding to $m=0$ modes ($k_y=0$). For reference some curves are also plotted that approximate quite well some of the spectra profiles in SLAB (see Figs.\ref{en_spectra_SLAB} and \ref{en_spectra_NEW_SLAB}) and which behave as: $1/k_\bot^3$ (dashed dark red line), $1/k_\bot^4$ (dash-dotted dark yellow line) and $1/k_\bot$ (dotted dark green line) for $E_{\tilde{p}}$ (left) and $1/k_\bot^{1.5}$ (dashed dark red line) and $1/k_\bot^2$ (dash-dotted dark yellow line) for $E_\phi$ (right).}
\label{en_spectra_NEW_shear}
\end{figure}

2D energy spectra are shown in Fig.\ref{en_spectra_NEW_shear} and they have generically similar profiles but stronger anisotropy with respect to those for fully periodic BC in cases a) (red), b) (green) and c) (blue). This can be seen by comparing the amplification of $m=0$ modes (with black edges) with respect to other modes at same $k_\bot$ values, that is here more intense than in Fig.\ref{en_spectra_shear}. In case d) (yellow), the  $E_{\tilde{p}}$ energy spectrum is stiffer with respect to the reference curve (dark yellow) and it resembles the profile for a decaying turbulence scenario. In fact, now turbulence gets reduced, differently from the case with fully periodic BC, and it is not able to sustain the pressure profile that decays in time for both cases c) and d). Since such dissipative phase is diffusivity dominated the fraction of $n=0$ modes stays almost constant.  

This is confirmed by the flux analysis for $\tilde{p}$, while that for $\phi$ provides similar results as for fully periodic BC with 2D and toroidal transitions and the linear contributions having similar magnitudes and all contributing to the relaxation to the final energy spectra. 

Therefore, implementing vanishing radial Dirichlet BC has a direct impact on the degree of toroidal and poloidal symmetry of the final field configurations, especially for $\tilde{p}$, on the anisotropy of energy spectra and also on the fate of the saturated $\tilde{p}$ profile, if stationary or time-decaying.  

\section{Conclusions}\label{sec4}
The simulations for the system of equations (\ref{ES}) describing drift turbulence in the edge of a DTT-like tokamak plasma have been presented. We outlined the rich phenomenology of the resulting 2D energy spectra whose profiles are affected by the choice of model parameters (background pressure gradients and diffusivity), the magnetic geometry (the expression of the poloidal magnetic component in addition to the toroidal one), the radial BC (periodic vs vanishing Dirichlet ones).    

The choice of the model parameters determines the relevance of different mechanisms affecting the final field configuration and, in particular, the degree of toroidal symmetry. The magnitude of background pressure gradient, modeled by the inverse of $\ell_0$, determines the amount of $\tilde{p}$ free-energy injected into the system. Such free-energy is subject to a direct toroidal cascade, thus by increasing background pressure gradient, {\it i.e.} reducing $\ell_0$, one gets more toroidally asymmetric $n\neq 0$ modes. However, the cascade decreases by increasing diffusivity $\chi_\bot$ from classical to neoclassical values, that provides major dissipation at large perpendicular scales and minor energy available for the direct toroidal cascade and subsequent transfer to $E_\phi$. We discussed two cases for $\ell_0$: weak, $\ell_0=10cm$, and strong $\ell_0=2cm$ background pressure gradients. For strong gradients, the cases with classical and neoclassical diffusivities differ just quantitatively, with the former showing more toroidal symmetry and more narrow spectra with respect to the latter. For weak gradients, we found in all the considered magnetic configurations a substantial difference between classical and neoclassical $\chi_\bot$: in the latter turbulence is time-decreasing, a sheared radial electric field develops and field profiles show maximal degree of toroidal symmetry. The classical case is qualitatively similar to those with strong gradients and in SLAB we recover the spectral profile $\sim 1/k_\bot^2$ of 2D inviscid Euler equation derived in the reduced model of Ref.\cite{Montani:22}.  The crucial role of the diffusivity parameter for the emerging turbulent scenario and field spectral features suggests that such parameter should be chosen with great caution in simulations and maybe it is worth planning some diagnostics to measure directly some (pressure and/or potentials) spectral profiles in order to validate our model assumptions (or the findings of consistent models for diffusion such as the $\kappa-\epsilon$ model \cite{Bufferand:21}).   
     
The addition of a poloidal magnetic component to the toroidal one produced strongly anisotropic 2D spectra with more energy for small $m$s. This is ascribed to the poloidal component of the drift term, that provides an effective dissipative contribution vanishing at $m=0$ and increasing with $m$. The resulting spectra for a magnetic configuration that is radially sheared between SLAB and one with a constant poloidal component exhibit anisotropic features due to the poloidal drift and average profiles that roughly follows those in SLAB. The simulations for the X-point magnetic geometry will be presented in the companion paper \cite{Xpoint} and the achievements of this work will be used to interpret the results.    
   
A further source of spectral anisotropy is here given by the choice of vanishing radial Dirichlet BC, instead of fully periodic ones, that breaks the formal symmetry between radial and poloidal coordinates, resulting in a mild enhancement of $m=0$ modes, mostly for $\tilde{p}$, and in turbulence quench. Understanding the role of BC and their spectral implications could be relevant to solve the discrepancy between global simulations and experiments in the divertor region, where sheat BC are imposed at the divertor plates. For this reason, further developments will deal with the implementation of sheat BC, via the penalization technique (as in GDB \cite{Zhu:18} and GRILLIX \cite{Stegmeier:19}) or by mimicking the procedure adopted in Refs.\cite{To:20,Wang:20} to implement nontrivial BC with FFT. 

The analysis of fluxes clearly pointed out that toroidal cascades and 2D transitions cannot be treated independently as they are strongly correlated and they provide contributions of the same order that sum up with linear ones to drive the system towards the saturated spectral profiles. This means that 3D simulations are inescapable in order to account for the dynamical features of drift turbulence, as argued by the pioneering work \cite{Biskamp:95}. 

The inclusion of other drift effects, in particular the curvature contribution accounting for interchange instability, of the diamagnetic velocity, whose role in triggering oscillations is outlined in Ref.\cite{Cianfrani:22}, and the investigation of magnetic fluctuations via the solution of the electro-magnetic system of equations (\ref{EM}) will be pursued in the follow-up analyses. The aim is to understand the relevance of the present achievements for a complete description of edge tokamak turbulence in view of a comparison with fully-poloidal simulations that could help distinguishing spatially local from global effects.
 

\end{document}